\documentclass[a4paper,fleqn,usenatbib]{mnras}

\usepackage[T1]{fontenc}
\usepackage{ae,aecompl}

\usepackage{graphicx}
\usepackage{gensymb}
\usepackage{amsmath}	
\usepackage{amssymb}	
\usepackage{txfonts}
\usepackage{float}
\usepackage{amsfonts}
\usepackage[utf8]{inputenc}

\title[MRI driven wind cycles in local disc models]{Magnetorotationally driven wind cycles in local disc models}

\author[]{
A. Riols,$^{1}$, G. I. Ogilvie $^{1}$, H. Latter $^{1}$ and J. P. Ross $^{1}$
\\
$^{1}$Department of Applied Mathematics and Theoretical Physics, University of Cambridge, Centre for Mathematical Sciences, \\
Wilberforce Road, Cambridge CB3 0WA, United Kingdom. 
}

\date{Accepted XXX. Received YYY; in original form ZZZ}

\pubyear{2016}

\begin{document}
\label{firstpage}
\pagerange{\pageref{firstpage}--\pageref{lastpage}}
\maketitle
\begin{abstract}
Jets, from the protostellar to the AGN context, have been extensively studied
but their
connection to the turbulent dynamics of the underlying accretion disc is
poorly understood. 
 Following a similar approach to \citet{Lesur2013}, we examine the
 role of the 
magnetorotational instability (MRI) in the production and acceleration 
of outflows from discs. 
Via a suite of one-dimensional shearing-box
simulations of stratified discs
we show that
magneto-centrifugal winds exhibit cyclic activity with a period of $10-20\, \Omega^{-1}$, a few times the orbital period. 
{The cycle seems to be more vigorous for strong vertical field}; it is robust to the variation of relevant parameters and independent of numerical details. The convergence of these solutions (in particular the mass loss rate)  with vertical box size is also studied.
By considering a sequence of magnetohydrostatic equilibria and their
stability, the periodic activity may be understood as the succession
of the following phases: 
(a) a dominant MRI channel mode, (b) strong magnetic
field generation, (c) consequent wind launching, and ultimately (d)
vertical expulsion of the excess magnetic field by the expanding and accelerating gas associated with the wind.
 We discuss potential connections between this behaviour and
observed time-variability in disk-jet systems.

\end{abstract}

\begin{keywords}
accretion discs -- MHD winds -- instabilities
\end{keywords}



\section{Introduction}

Observations of accretion discs around T-Tauri stars, in X-ray
binaries, 
or on the larger scales of Active Galactic Nuclei (AGN), have revealed 
the existence of supersonic outflows and collimated jets
\citep{bally07,fender03,schwartz07}. 
Sequences of geyser-like eruptions occurring at 
quasi-regular intervals have been observed in protostellar systems
\citep{zinnecker98,hartigan07}, while 
relativistic jets emitted by AGN also exhibit rich
time-variability, 
characterized by knotty structures revealed in particular by the CHANDRA X-ray imaging of M87 \citep{harris09}.
From a theoretical perspective, these powerful ejections (steady or
not) are thought to be launched from accretion discs and possibly
related to the magneto-centrifugal effect as proposed by
\citet{blandford82}. In this model, a large-scale poloidal magnetic
field, anchored in the disc and corotating with it, is believed to
accelerate the gas. When both centrifugal and gravitational forces are
taken into account, particles are accelerated along the field lines
when their inclination from the rotation axis is more than
30\degree. Magnetic field lines are able to extract angular momentum
from the disc and therefore might take part in the accretion
process. Detailed models of magnetically driven jets
\citep{lovelace91,pelletier92} and simulations of magneto-centrifugal
ejection \citep{romanova97, ouyed99, krasno97} have been
produced. However, they usually assume that the global magnetic
configuration is supported by the accretion disc, neglecting the
back-reaction of the disc dynamics on the large-scale magnetic field
and vice versa. Also, most of them treat the disc as a boundary
surface that loads mass on to the magnetic field lines at a given
rate, but in fact the conditions for gas to escape from the disc are still not entirely understood. 

More recent work has investigated the effect of the vertical structure
of the disc on the outflow. In the case of a strong poloidal field
that threads the disc with a gas to magnetic pressure ratio $\beta_0
\lesssim 1$ at the midplane, it has been shown that significant
thermal assistance is required to overcome the potential barrier
between the surface of the disc and the first critical point, namely
the slow magnetosonic point \citep{ogilvie97,ogilvie98}. In a
symmetric disc, this barrier is due to the bending of the field lines
close to the midplane. \citet{ogilvie01} showed that, in the absence
of an additional source of energy, the mass-loss rate decreases very
sharply with increasing field strength and is maximized at an
inclination of approximately $45\degree$. \citet{ogilvie12}
demonstrated that quasi-steady outflows in this regime can be
simulated in a local 1D shearing-box model as this contains most of the relevant physical elements.\\
 
An interesting regime is the case of weaker fields for which $\beta_0
\gtrsim 1$ at the midplane. The situation is then altered because
the disc is subject to the magnetorotational instability
 \citep[MRI;][]{velikhov59,chandra60,balbus91}. The interaction
 between the MRI dynamics and the wind outflow has been studied in the
 last few years but still remains debated. Simulations of 3D MRI
 turbulence in a shearing box, with a very weak net vertical magnetic
 field ($\beta_0$ from $10^2$ to $10^5$) have shown that unsteady
 outflows can be produced, starting about two scale heights above the
 midplane \citep{suzuki09,fromang13, bai13b}. The occurrence of
 transient and local outflows appears to be a consequence of the
 nonlinear development of the MRI in stratified discs, even if the
 connection between local outflows and the observed large-scale jets
 remains unclear. \citet{Lesur2013} showed in a 1D model that a wind
 can be produced even when starting with a purely vertical field, as
 the field lines are bent during the growth of MRI channel modes.
 Magnetic pressure gradients enhanced by MRI modes are thought to
 play an  important role in the launching process as they help to
 overcome the potential barrier, while far from the disc the magneto-centrifugal effect accelerates the gas. \\

Although steady or quasi-steady outflows seem to appear in the
MRI-unstable weak-field regime, their properties depend strongly on
the numerical boundary conditions of the simulation. In particular, the mass-loss rate decreases
with the height of the box
\citep{moll12,Lesur2013,fromang13}. Simulations that take into account
the disk's vertical structure and its symmetry also fail to obtain
steady jets with velocities exceeding the fast magnetosonic speed. The
fact that these solutions depend on the boundary conditions applied at
the top of the shearing box begs the question
of their connection with the external medium, and even casts doubt on
their existence in reality. Other important
issues, which are not addressed in the present paper, are
the sensitivity of MRI-driven outflows to non-ideal effects such as Ohmic
resistivity, the Hall effect and ambipolar diffusion \citep{bai13a,
  gressel15} and the problem of the radial transport of poloidal  
magnetic flux, in which MRI turbulence is thought to play a crucial role \citep{guilet13}. \\

In this paper, we aim to extend previous numerical studies showing that the MRI can excite unsteady winds in the disc atmosphere and quasi-periodic outbursts
\citep{Lesur2013,fromang13}, by simulating a regime closer to the marginal stability boundary of the MRI ($1 < \beta_0 < 10^{4}$) and investigating in more detail the nonlinear 
interaction between the instability and these outflows.  We show that
outburst cycles are a robust feature in 1D vertically stratified
simulations and depend on the strength of the imposed vertical field. 
We present an interpretation of the phenomena which treats the cycle
as a sequence of quasi-steady MHD equilibria (which we compute), 
with the engine driving
the system from state to state being either MRI channel modes, or the
loss of magnetic flux via the ensuing outflow. The cycle can then be
broken down into the following successive phases: 
(a) the emergence of a dominant MRI
channel mode, (b) the generation of strong horizontal magnetic
field by the channel, (c) suppression of the MRI by the
field and the
concurrent launching of a wind, and (d)
vertical expulsion of
the excess magnetic field by the wind and a return to the initial laminar weak
field state. This behaviour could potentially be connected to the
variability both in accretion disk and jet emission, but is probably
limited to regions of the disk where the MRI is not far from
marginality, i.e.\ the magnetic flux or non-ideal MHD effects are
locally strong. 

The structure of this paper is as follows.
In section
2, we describe the equations and justify the use of a one-dimensional
local model. In section 3, we explain the choice of our numerical
setup and in particular the tall boxes used for this study. In section
4, we present the phenomenology of time-periodic solutions found in the MRI
unstable regime and show that they exist in a large range of parameter
values. In section 5, we propose a mechanism that might explain the
cyclic process. We calculate the MRI stability of magnetohydrostatic
disc equilibria, and connect the results with the nonlinear dynamics
of periodic outflows. We finally discuss in section 6 the relevance of
our model for accretion discs  
and the possible implications of our cyclic solutions for astrophysical applications.
\section{Model and equations}
\label{model}
We use the Cartesian local shearing-sheet description of differentially rotating flows \citep{goldreich65}, whereby the radial, azimuthal and vertical directions are denoted respectively $x$, $y$ and $z$. The axisymmetric differential rotation is approximated locally by a uniform rotation $\mathbf{\Omega}=\Omega \, \mathbf{e}_z$ plus a linear shear flow  $\mathbf{U}=-Sx\,  \mathbf{e}_y$  with  $S=(3/2)\,\Omega$ for a Keplerian disc. We assume that perturbations are axisymmetric, which means that they do not depend on $y$ in the local framework. As the local approximation has a translational symmetry in $x$, it is possible to study magnetohydrodynamic (MHD) solutions that depend only on $z$. This assumption does not reproduce the complex physics occurring in a realistic disc but might capture the essential mechanisms operating in the wind-launching process. We also note that some 3D studies have found that the dynamics reduces naturally to a laminar 1D form \citep{bai13}.

For simplicity we also assume that the gas is isothermal so that its
pressure and density are related by $P=\rho c_s^2$ with $c_s$ the
uniform sound speed. The system of equations governing the evolution
of the fluid density $\rho$, velocity $\mathbf{v}$ and magnetic field
$\mathbf{B}$, in the 1D approximation, is \citep{ogilvie12}
\begin{equation}
\dfrac{\partial \rho} {\partial t}+\dfrac{\partial (\rho v_z)}{\partial z}=\varsigma(z,t),
\label{eq_rho}
\end{equation} 
\begin{equation}
\rho\left(\dfrac{\partial v_x}{\partial t}+v_z\dfrac{\partial v_x} {\partial z}-2\Omega v_y\right)=\dfrac{\partial}{\partial z}\left(\dfrac{B_xB_z}{\mu_0}+\rho\nu\dfrac{\partial v_x}{\partial z}\right),
\label{eq_ux}
\end{equation} 
\begin{equation}
\label{eq_uy}
\rho\left(\dfrac{\partial v_y}{\partial t}+v_z\dfrac{\partial v_y} {\partial z}+\dfrac{1}{2}\Omega v_x\right)=\dfrac{\partial}{\partial z}\left(\dfrac{B_yB_z}{\mu_0}+\rho\nu\dfrac{\partial v_y}{\partial z}\right),
\end{equation} 
\begin{equation}
\label{eq_uz}
\rho\left(\dfrac{\partial v_z}{\partial t}+v_z\dfrac{\partial v_z} {\partial z}\right)=\rho g_z+\dfrac{\partial}{\partial z}\left(-c_s^2\rho -\dfrac{B_x^2+B_y^2}{2\mu_0}+\dfrac{4}{3}\rho\nu\dfrac{\partial v_z}{\partial z}\right),
\end{equation} 
\begin{equation}
\dfrac{\partial B_x} {\partial t}=\dfrac{\partial}{\partial z}\left(v_xB_z-v_zB_x\right)+\eta \dfrac{\partial^2 B_x}{\partial z^2} ,
\label{eq_bx}
\end{equation} 
\begin{equation}
\dfrac{\partial B_y} {\partial t}=-\dfrac{3}{2}\Omega B_x+\dfrac{\partial}{\partial z}\left(v_yB_z-v_zB_y\right)+\eta \dfrac{\partial^2 B_y}{\partial z^2},
\label{eq_by}
\end{equation}
\begin{equation}
\dfrac{\partial B_z} {\partial t}=0,    \quad \quad  \dfrac{\partial B_z} {\partial z}=0.
\label{eq_bz}
\end{equation}
where $\nu$ and $\eta$ are respectively the kinematic viscosity and magnetic diffusivity. It is possible to define $H=c_s/\Omega$ as the standard hydrostatic scaleheight of the disc. This quantity will be considered as our unit of length, while  $\Omega^{-1}$ is our unit of time. The source term $\varsigma(z,t)$ is an artificial mass injection that mimics the mass replenishment that could occur in a realistic disc when curvature effects and radial transport are taken into account. To obtain steady or periodic solutions of equations (\ref{eq_rho})--(\ref{eq_bz}), this term has to compensate the loss of mass being ejected by MHD winds. The way mass is replenished is explained in more detail in section \ref{bc_replenish} and discussed in section \ref{rob_replenish}.\\


The gravity term $g_z$  reduces to $-\Omega^2\,z$ in the standard local approximation. However, this approximation is only appropriate when the vertical scales of interest are small compared to the disc radius. For the reasons explained in section \ref{box_size_res}, wind solutions are better studied in the limit $z\gg H$ with $z$ comparable to the radial extent of the disc. It is then necessary to introduce an additional length scale $r_0$, which can be regarded as the radius of the reference orbit on which the local model is constructed, i.e.\ the distance of the origin of the shearing box from the central object. For $z \gtrsim r_0$ the vertical component of the central gravity starts to decrease and tends to zero further away. Assuming that the disc can be defined by its local thickness or by its dimensionless aspect ratio
\begin{equation}
\delta=\frac{H(r)}{r},
\end{equation}
the radius $r_0$ can be expressed in our unit system as $H/\delta$. The modified vertical gravity taking into account the finite distance from the central object is then
\begin{equation}
\label{eq_gravity}
g_z=-\dfrac{GMz}{(r_0^2+z^2)^{3/2}} = -\Omega^2H \dfrac{\hat{z}}{(1+\delta^2 \hat{z}^2)^{3/2}}
\end{equation}
where $\hat{z}=z/H$. Similar corrections to the vertical gravity have
been used by previus authors when modelling accretion discs
\citep{matsusaki97} or galaxies \citep{kuijen89}. For $\delta=0$, we
retrieve the standard gravity of the local thin disc
approximation. The typical $\delta$ for an astrophyscial
protoplanetary disc, inferred from recent observations of T-Tauri
stars, is between 0.03 and 0.2  \citep{andrews09,Grafe13}. 
Note that this gravity law is not fully consistent with the local
approximation of Keplerian flows as the radial gravity force also changes
with altitude. However, in order to simplify the problem and still use
the one-dimensional model, we will assume that 
this variation of the radial gravity is unimportant for the dynamical
flow we are studying. 

\section{Numerical setup and parameters}

\subsection{Numerical Code}
We used the shearing-box approach to simulate locally the one-dimensional disc atmosphere. As the flow is compressible and possibly gives birth to a variety of shocks, we used the PLUTO code \citep{mignone07}, a finite-volume method with a Godunov scheme that integrates equations (\ref{eq_rho})--(\ref{eq_bz}) in their conservative form. The fluxes are computed with the HLLD Riemann solver but we checked that solutions are independent of the choice of solver. The time-integration is achieved by a Runge--Kutta method of third order. We emphasize that 1D simulations can be time-consuming, especially in regimes or locations where the typical Alfv\'en speed $V_{Az}=B_z/\sqrt{\mu_0\rho}$ is large compared to $c_s$. \\

\subsection{Boundary conditions and mass replenishment}
\label{bc_replenish}
As in \citet{ogilvie12} and \citet{Lesur2013}, we impose symmetry with respect to the disc midplane. Simulations are then restricted to the upper half of the disc, as the lower half is obtained by taking $\rho(-z)=\rho(z)$, $v_x(-z)=v_x(z)$, $v_y(-z)=v_y(z)$, $v_z(-z)=-v_z(z)$, $B_x(-z)=-B_x(z)$ and $B_y(-z)=-B_y(z)$. This symmetry implies that the poloidal magnetic field is vertical in the midplane, i.e $B_x(z=0)=0$. With such symmetry, we guarantee that field lines bend in the disc midplane, which makes the disc structure more stable.\\

For the upper boundary $z=L_z$, we use standard outflow boundary conditions for the velocity field but impose hydrostatic balance in the ghost cells for the density, in a manner similar to \citet{simon11}. In this way, we reduce significantly the excitation of artificial waves near the boundary. As explained in more detail in section \ref{rob_bc}, various boundary conditions for the magnetic field have been used. A first possible way (Type I boundary) is to fix $B_x$ (and therefore the inclination of the poloidal field)
and impose zero gradient for $B_y$ as employed by \citet{Lesur2013}. Another is to make a linear extrapolation of $B_x$ and $B_y$ into the ghost cells (Type II boundary). Unless it is explicitly mentioned, all simulations have been run with the first type of boundary condition. \\

In order to reproduce the mass replenishment due to the radial redistribution of material, we inject mass near the midplane at each numerical time step. The source term in the mass conservation equation (\ref{eq_rho}) is similar to the one prescribed by \citet{Lesur2013},
\begin{equation}
\varsigma(z,t)=\dfrac{2\dot{m_i}(t)}{\sqrt{2 \pi} z_i}\exp{\left(-\dfrac{z^2}{2z_i^2}\right)}, 
\label{injection}
\end{equation}
where $\dot{m_i}(t)$ is the mass injection rate (into the upper half of the disc) and $z_i$ is a free parameter that corresponds to the altitude below which most mass is replenished. There are two different ways to replenish the mass. The first one is to maintain a constant 
disc surface density $\Sigma=2{\int_0}^{L_z} \rho \, dz$ in
time. Then, by computing the amount of mass that leaves the box, it is
straightforward how much must be injected each time step. Perhaps a more physically relevant method is to inject mass at a constant rate $\dot{m_i}$. This approach is probably more realistic but suffers from the fact that the mass contained in the computational domain can vary in time (periodically if the solution of interest is a cycle). We use preferentially the first method and will see in section \ref{rob_replenish} how results are affected by choosing the other method. Note that mass is injected without any velocity, although it leaves the box with a non-zero velocity, which implies that the total momentum in the box is not conserved. In fact the loss of horizontal momentum from the box (which can also occur through magnetic stresses acting at the upper boundary) drives a mean horizontal flow, which can be interpreted as an accretion flow together with a small departure from Keplerian rotation.

\subsection{Box size and resolution}
\label{box_size_res}
As pointed out by \citet{fromang13} and \citet{Lesur2013}, wind properties and mass-loss rates obtained from shearing-box simulations depend on the box size $L_z$ and the upper boundary conditions. This might be a consequence of the fact that critical points, such as the Alfv\'en point or the fast magnetosonic point, often lie outside the computational domain  [definitions of these points are given by \citet{ogilvie12}]. Crossing the Alfv\'en point for $\beta_0 \gtrsim 1$ is possible by having a tall box with $L_z\gg H$. However, crossing the fast magnetosonic point is more delicate. Previous numerical studies in 1D or 2D models have failed to compute realistic steady solutions that cross this critical point \citep{Lesur2013}, unless the temperature of the atmosphere is strongly enhanced \citep{ferreira04} or the vertical symmetry of the disc is broken \citep{vlahakis00}. In our case, we did not find any solutions that pass through the fast point. Therefore the boundary conditions and the box size $L_z$ may still influence the numerical solutions, although we found surprisingly that several wind properties seem to converge when a large vertical height is reached (see section \ref{rob_bc}).  \\

Another drawback inherent to previous shearing-box simulations is that $g_z$ increases linearly with $z$ in the upper disc atmosphere, corresponding to a gravitational potential well of limitless depth; this might lead to solutions that are artificial or dependent on boundary conditions. In particular it is possible to show that any steady solutions with $g_z\nrightarrow 0$ as $z\to\infty$ cannot exist in our setup (see appendix \ref{appendixA}). This suggests that the behaviour of solutions at high altitude $z$ might be better assessed by using the modified gravity of equation~(\ref{eq_gravity}) provided that $L_z/H \gtrsim 1/\delta$.\\

We then chose a suitable box size $L_z=70\, H$ that satisfies the
different conditions  discussed before (except the crossing of the
fast point) for the typical range of parameters we used.  {Note
  that the main limitation of the shearing box model when considering
  such large $L_z$ is that magnetic field lines in the upper
  atmosphere are anchored to radii far from $r_0$. In
  reality, these radii will rotate at different periods from those
  implied by the uniform shear of the local approximation, suggesting that our model is unlikely
  to be consistent with the global disc equilibrium. However,
  given the robustness of the phenomena we uncover, we doubt that this
  inconsistency impacts on our results unduly}.  
 A second problem is that the MHD approximation may fail in such tall
 boxes as the mean free path in the upper atmosphere may be too long. 
 In actual fact this is rarely the case since outflows supply enough mass in
 this region: the typical 
 density measured at the upper boundary never drops below $10^{-4}$ of the midplane density. \\
\begin{figure*}
\centering
\includegraphics[width=\textwidth]{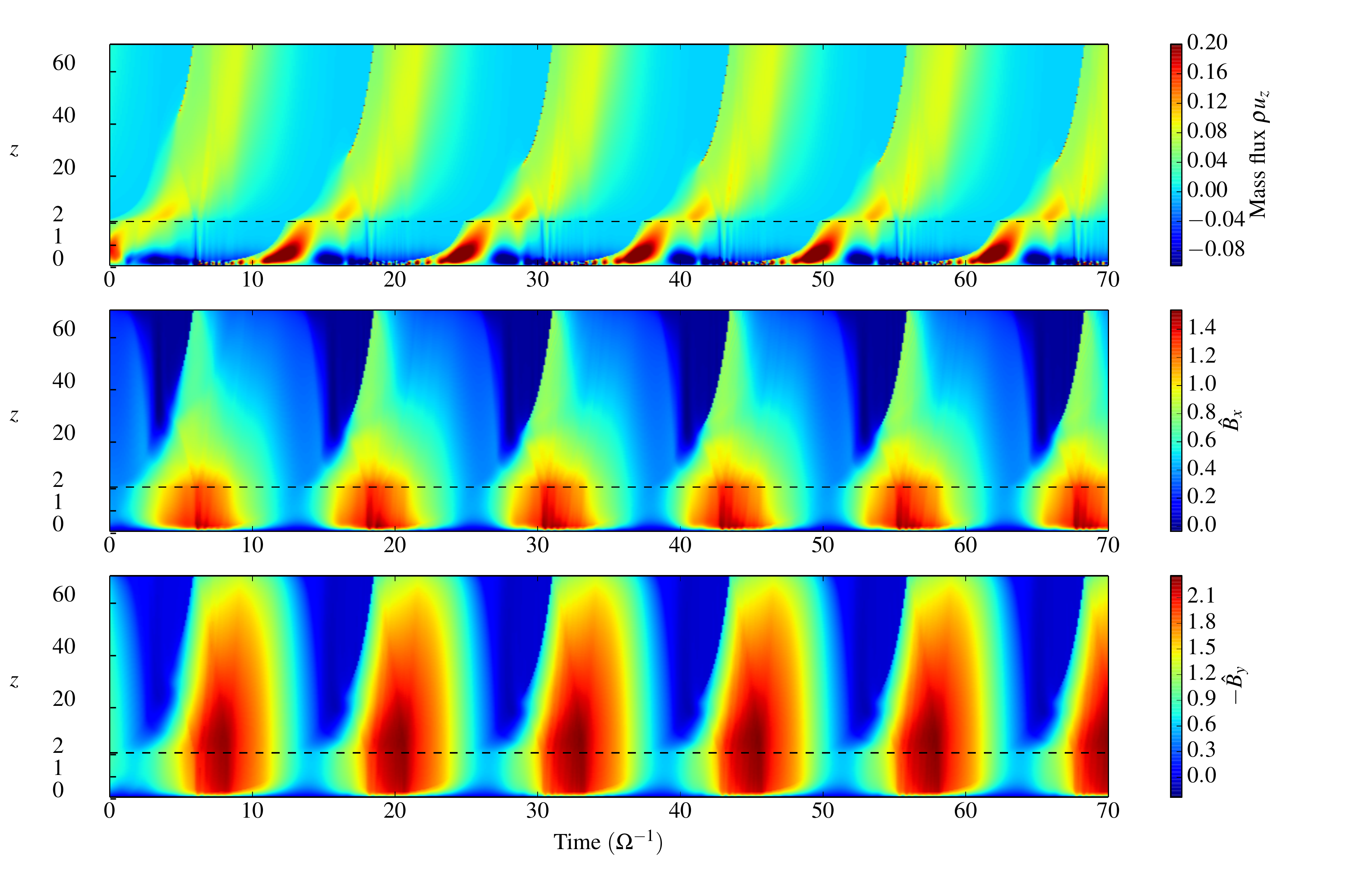}
 \caption{Space-time diagrams of a periodic ideal MHD wind solution computed for $\hat{B}_{z}=0.54$, $\mathbf{\delta=0.033}$  and $z_i=0.5$. Top, middle and bottom colourmaps show respectively the mass flux $\rho v_z$, radial  ($\hat{B}_x$) and  azimuthal ($\hat{B}_y$) magnetic field. The black/dashed line at $z=2$ corresponds to the separation between the refined and the coarse grid patches. It is also used conveniently to dissociate the `lower' atmosphere ($z\lesssim 2$, including the midplane) from the `upper' atmosphere ($z\gtrsim 2$). }
\label{fig_cycle1}
\end{figure*}
Our numerical grid is a combination of two uniform patches with  $N_z=1000$ in the domain $0 \leq z \leq 2 H$ and $N_z=4000$  for $2 H \leq z \leq 70 H$. The reason for using a finer grid in the midplane region is that the disc scaleheight can reach $\sim 0.1 H$ when the material is subject to strong magnetic compression.  In addition, small-scale structures like pressure waves arise more naturally near the midplane.
For comparison, we also used a uniform grid with resolution $N_z=4000$ and checked that the solutions do not differ significantly.\\
\subsection{Physical parameters}

In all simulations (except the one with constant $\dot{m_i}$ shown in Fig.~\ref{fig_cycle_bc} \& \ref{fig_cycle_mi}), we fixed the surface density $\Sigma$ to be equivalent to the column mass of a hydrostatic disc with midplane density $\rho_0=1$, which sets our unit of mass.  In the one-dimensional approximation, the vertical magnetic field $B_z$ is independent of $z$ and $t$. A suitable dimensionless parameter for this problem is then
\begin{equation}
\label{def_bzbar}
\hat{B}_{z}=\dfrac{B_z}{\sqrt{\mu_0\Sigma c_s\Omega}}.
\end{equation}
We focus in this paper on the dynamics in the range
$\hat{B}_{z}\lesssim 1$,  which translates to an average plasma $\beta_z\gtrsim1$, defined as: 
\begin{equation}
\beta_z=\dfrac{1}{\hat{B}_z^2}
\end{equation}

which remains close to the midplane $\beta_0$ if the disc is not significantly compressed. We expect the MRI to be active in this regime, giving rise
to significant outflows.
\\

We assumed ideal MHD ($\eta=0$), except in section \ref{rob_dissipation} where the role of ohmic diffusion is studied. We introduced a very tiny kinematic viscosity in the midplane, $\nu=10^{-5}$.  Without any substantial explicit viscosity near the upper boundary, strong numerical instabilities appear, in particular for $\hat{B}_{z}\simeq 0.5$, for which outflows can be 80 times faster than the sound speed. In order to avoid these instabilities, we forced $\nu$ to be inversely proportional to density at large $z$.  Actually this prescription might be physically relevant, at least in the isothermal case. Indeed, as long as the plasma remains collisional, $\nu$ can be estimated as the product of the thermal velocity and the mean free path, which scales as $1/\rho$ \citep{maxwell66}. Whether this prescription is realistic or not, we checked that viscosity plays a minor role in the results. Viscous forces in the upper atmosphere remain very small compared to inertial forces, while changing the value of $\nu$ at the base does not strongly modify the solutions, as long as numerical instabilities are damped. 

\section{Time-periodic wind solutions}
\label{periodic_sol}

\subsection{Initial conditions and capturing a cycle}

The set of initial conditions that converge towards a periodic
solution is not necessarily easy to guess. These solutions do not
appear spontaneously in small boxes and their study requires the use
of large $L_z$. However, starting from an MRI-unstable hydrostatic equilibrium
in very tall boxes with small $\delta$ is numerically unachievable because of excessively high Alfv\'en speeds.     
We therefore began with a small box of size $L_z=6$ and
$\delta=0$. The initial condition consists of a hydrostatic disc
embedded in a vertical field (originally we chose
$\hat{B}_{z}=0.5$). Small perturbations were introduced and were
amplified by the MRI instability. The nonlinear state evolved rapidly
into a steady outflow, as described by \citet{Lesur2013}. We then
progressively continued this steady solution by increasing the box
size and found that it bifurcates to a periodic outflow at $L_z\simeq
15$. As shown later in section \ref{rob}, cyclic wind solutions are
fairly robust to any change in the numerical or physical parameters
and we believe they have a physical reason to exist. Unlike the steady
outflow obtained for $L_z \lesssim 15$, the properties of these cyclic
solutions seem to converge as the box size is increased (see section
\ref{rob_bc}). 

 Another method to obtain periodic outflows is to start from the
 perturbed hydrostatic equilibrium in a very tall box by taking
 $\delta\simeq 1$. In that case, the density drops moderately at large
 altitude  $z$  so that the Alfv\'en speed (and therefore the
 computational time)  do not diverge too rapidly with $z$. By means of
 parameter continuation, 
$\delta$ is finally decreased to obtain more realistic solutions.

\subsection{Phenomenology of wind cycles}
\label{cycle_pheno}
We present in this paragraph an example of a cyclic solution obtained
for  $\hat{B}_{z}=0.54$ ($\beta_z=3.4$), {$\mathbf{\delta=0.033}$} and $\hat{z}_i=0.5$. Figure
\ref{fig_cycle1} shows the spatial variation and temporal evolution of
the magnetic components $\hat{B_x}$ and $\hat{B_y}$ (normalized in the same way as eq.~\eqref{def_bzbar}) and vertical mass flux during 70
$\Omega^{-1}$. The period of the cycle is roughly equal to $12.5\,
\Omega^{-1}$, i.e.\ two orbital periods. The dashed line at $z\sim 2H$
denotes the separation between the coarse grid and the fine
grid. Actually this virtual line remains very close to the sonic line
during the whole simulation. This is true for a uniform grid as well,
and thus is not an artefact of splitting the box into two domains. It
is convenient to  use this particular altitude as a separation between
the `lower atmosphere',  dominated by the disc dynamics, and the
`upper atmosphere'  dominated by the wind flow. Another important
altitude,  located below $z=H$, is where the toroidal current and
magnetic  pressure gradient change sign (going from positive to negative).\\

The first colourmap (Figure \ref{fig_cycle1}, top) shows that the cycle is made of two phases. One is quiescent with a very weak outflow, $\rho v_z \sim 0$, occurring between $t=0 $ and $t=10\,\Omega^{-1}$ near the midplane. The second consists of a violent ejection of gas starting around $t=10\,\Omega^{-1}$ in the lower atmosphere, with a mass flux $\rho v_z \sim 0.1$  measured at the upper boundary a few $\Omega^{-1}$ later. The second colourmap shows that, during the cycle, the amplitude of the radial field varies quite strongly in the lower atmosphere ($\hat{B}_x$ varies between 0.3 and 1.4) whereas it remains confined to smaller values in the upper atmosphere. The toroidal field varies at the same frequency with a small positive time lag, comparable to $\Omega^{-1}$. \\

\begin{figure}
\centering
\includegraphics[width=\columnwidth]{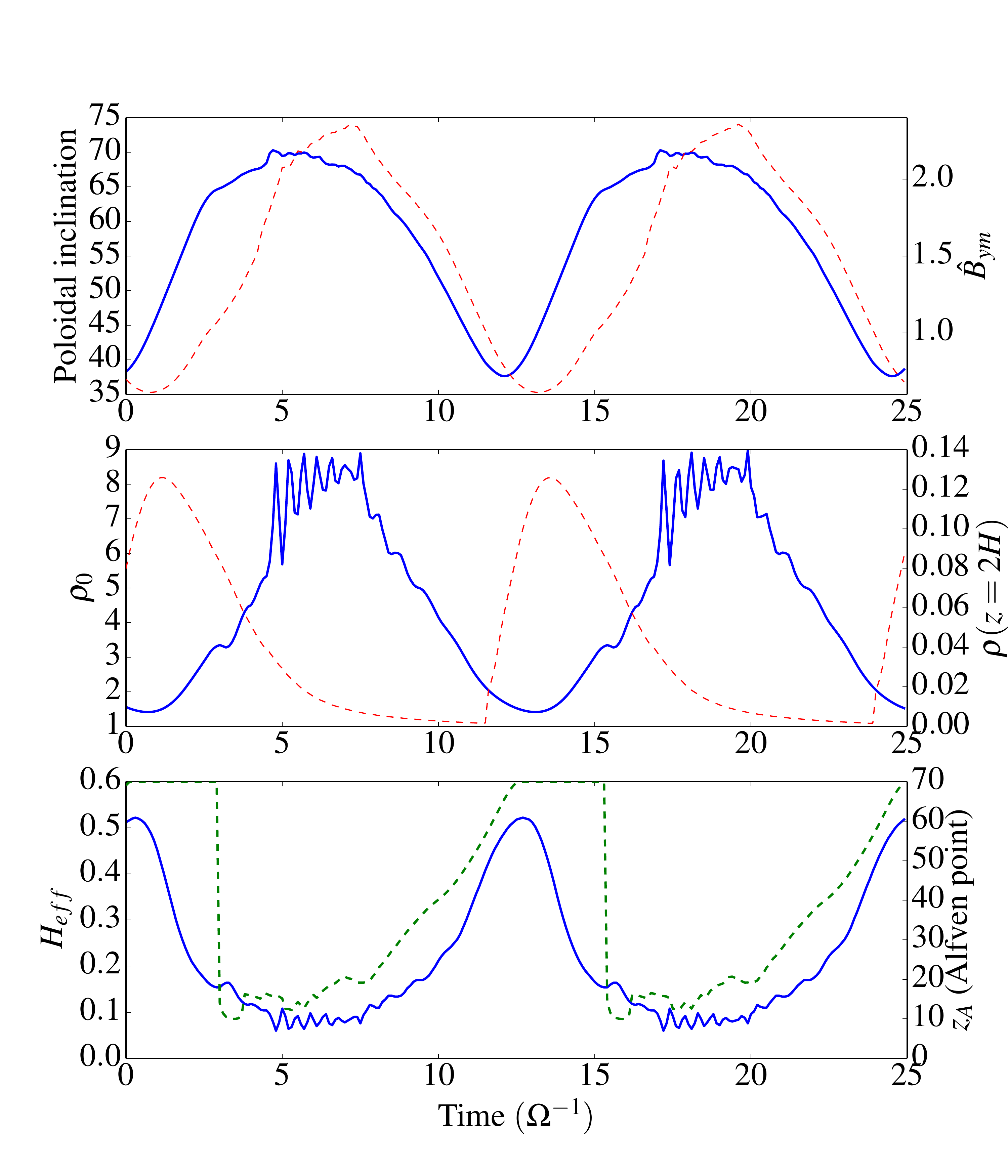}
\caption{Evolution of maximal inclination $i_m$, maximal normalized
  toroidal field $\hat{B}_{ym}$,  midplane density $\rho_0$, density
  at $z=2H$, effective disc scaleheight $H_{\text{eff}}$ and Alfv\'en point
  altitude $z_A$ during two periods of the cycle of
  Fig.~\ref{fig_cycle1}. The left labels correspond to the blue/solid
  curves whereas the right labels are for the red/dashed
  curves.}
\label{fig_cycle2}
\end{figure}

The outflow is initiated near the midplane (about one scale height above it) when $-\hat{B}_y$ is maximum and $\hat{B}_x$ starts to decrease. The inclination of the poloidal field at this time and for $z \sim 2H$ is around {60}\degree\,{with respect to the vertical axis}, allowing gas to be accelerated along field lines by the magneto-centrifugal effect. This inclination is smaller in the upper atmosphere (beyond the Alfv\'en point), but remains on average larger than 30\degree.   A strong toroidal field is produced in the lower atmosphere, well before the sonic point and Alfv\'en point are reached.  Angular momentum is then extracted from the disc by the toroidal field, allowing the flow to accrete (we checked that $v_x<0$ near the midplane). In the upper atmosphere and beyond the Alfv\'en point, its amplitude starts decreasing slowly with altitude, indicating that angular momentum is on the contrary transferred to the accelerating gas. Most of the phenomenology described here is reminiscent of the acceleration mechanism proposed by \citet{blandford82}. Note that the model is symmetric under reflection in $x$, so that another solution must exist in which the field lines bend the opposite way and angular momentum is added to the disc, resulting in a radial outflow. \\

To analyse the cycle in more detail, we plotted in Fig.~\ref{fig_cycle2} the evolution of several disc and outflow properties during two periods of the cycle. In particular, we show two dimensionless quantities that will be widely used in the next sections. They are the maximum poloidal inclination in $z$ 
\begin{equation}
i_m=\tan^{-1}\left[ \text{max}_z\,\left(\dfrac{B_x}{B_z}\right)\right],
\label{def_im}
\end{equation}
and the maximum normalized toroidal magnetic field along the same axis,
\begin{equation}
\hat{B}_{ym}=\text{max}_z\,\left(\dfrac{\vert B_y\vert}{\sqrt{\mu_0\Sigma c_s \Omega}}\right).
\label{def_Vay}
\end{equation}
At $t=0$, the magnetic variables $i_m$ and $\hat{B}_{ym}$ are close to
their minimum values. The disc is in a relatively expanded state as
its midplane density $\rho_0(t)$ is minimum while its effective
scaleheight $H_{\text{eff}}$, defined by $\rho (H_{\text{eff}}) =
\rho_0(t)e^{-0.5}$, is maximum (close to 0.5 $H$). The Alfv\'en point
is located far above the disc, in the upper atmosphere, indicating
that the wind is rather weak. Between  $t=0$ and  $t=5\,\Omega^{-1}$,
the radial and toroidal components of the magnetic field off the
midplane are
amplified. The magnetic pressure resulting from this amplification
rapidly compresses the disc material; its midplane density is multiplied by a
factor 3 while its effective scaleheight $H_{\text{eff}}$ is divided by a
similar factor.  At $t\simeq 5\, \Omega^{-1}, $  $\hat{B}_{ym}$
reaches a maximum while the poloidal field lines are strongly
inclined, with an angle of 70\degree. The midplane density and
$H_{\text{eff}}$ remain constant for a while, while the density near $z=2H$
keeps decreasing, suggesting that the gas in the disc is being
compressed by magnetic pressure. However a fraction of the material in
the disc atmosphere, located at the altitude where the current $J_y$
changes its sign, is expelled outward. Fig.~\ref{fig_cycle2} shows
that the Alfv\'en point drops suddenly and approaches the location of
the lower atmosphere, indicating that a wind is developing in the
lower atmosphere. Note that the vertical velocity (as inferred from
Fig.~\ref{fig_cycle1}) is rapidly oscillating during this phase and
before the emergence of the jet. Magnetosonic waves propagating upward
seem to be excited and are probably associated with the rapid
compression of the 
disc as its  height is oscillating with the same frequency. \\
\begin{figure*}
\centering
\includegraphics[width=\textwidth]{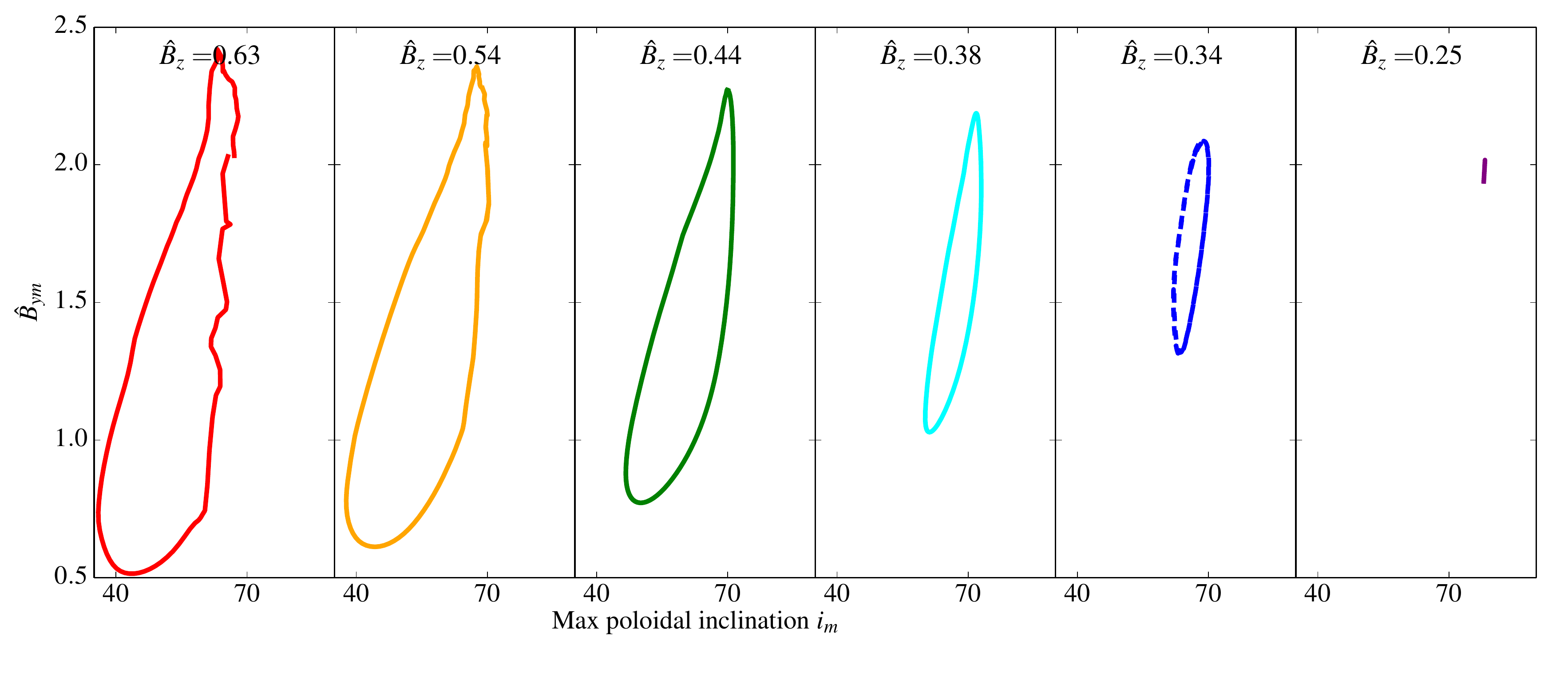}
 \caption{Left: Projection of the cyclic dynamics in a plane
   $(i_m,\hat{B}_{ym})$. Two periods are shown here, except for
   $\hat{B}_{z}=0.63$ for which only one period has been obtained. 
Solutions have been computed for a fixed $\Sigma$, $\delta=0.033$ and $z_i=0.5$ but with different $\hat{B}_{z}$ ($\beta_z$ varies from 2.5 to 16). The red/dashed curve corresponds to the largest $\hat{B}_{z}$ and the purple one to the smallest $\hat{B}_{z}$. Around $\hat{B}_{z}=0.25$, the solution bifurcates to a steady solution (fixed point) which is reduced to a point in this representation.}
\label{fig_cycle_Bz}
\end{figure*}
\begin{figure}
\centering
\includegraphics[width=\columnwidth]{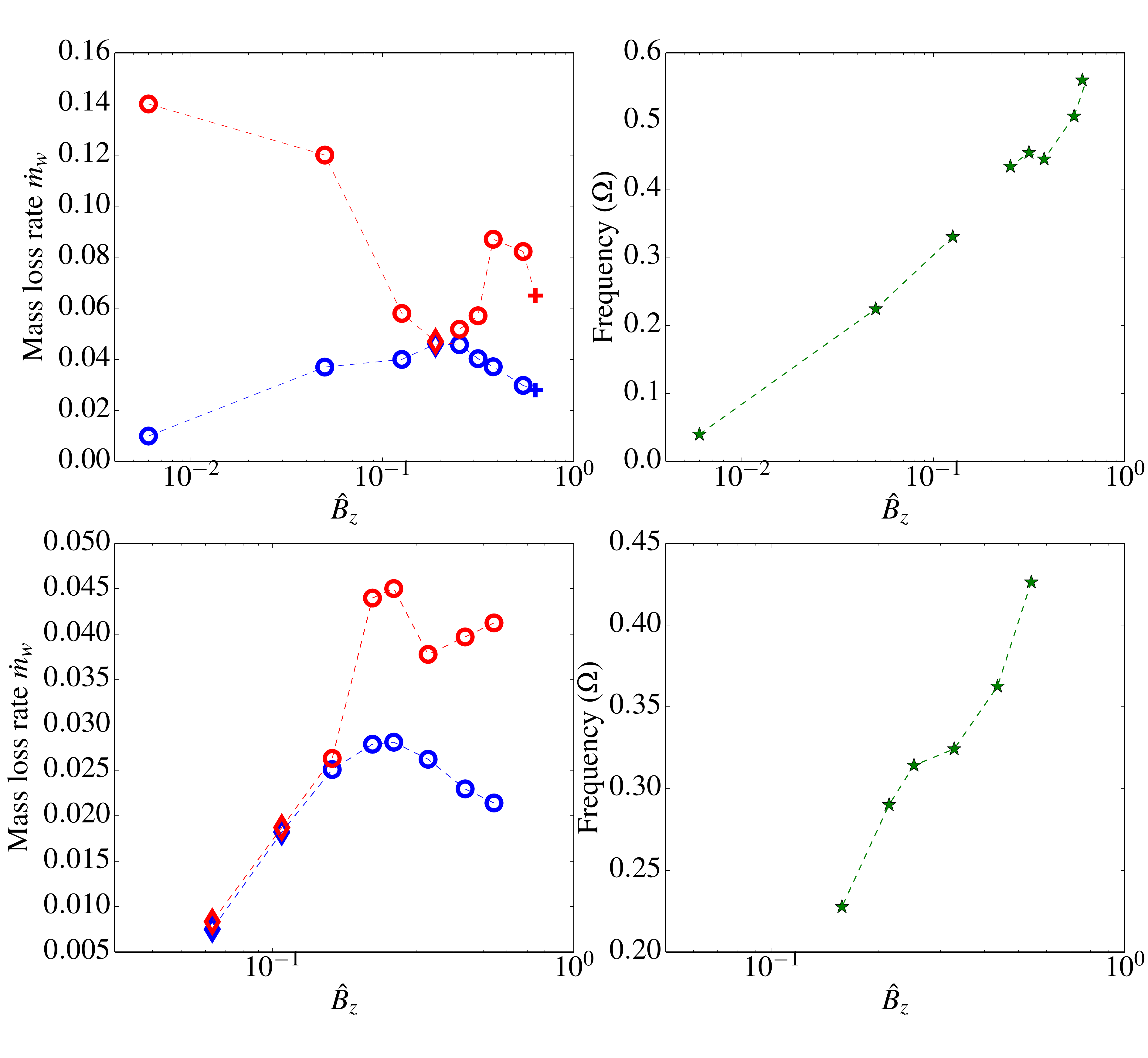}
 \caption{Left: Evolution of mean (blue) and maximum (red) mass-loss rate as a function of $\hat{B}_{z}$ ($\beta_z$ varies from 2.5 to $2.8\times 10^4$). The circle markers represent periodic solutions whereas diamond markers stand for steady solutions. Right: Cycle frequency as a function of $\hat{B}_{z}$. The top and bottom figures correspond respectively to $\delta=0.033$ ($z_i=0.5$) and $\delta=0$ ($z_i=1$).}
\label{fig_cycle_Bz2}
\end{figure}
Between $t=8\, \Omega^{-1}$ and $t=12.5\, \Omega^{-1}$, the wind accelerates rapidly and turns into a powerful outflow, manifested by the sudden increase of the density at $z=2H$. The jet finally ends when the mass in the lower atmosphere has been completely depleted. During the ejection phase, the amplitude of the magnetic field decreases and the poloidal field lines tend to their initial configuration with a small angle of $35\degree$. As the magnetic compression on the disc decreases, its typical height increases and comes back to its initial value.  At $t=12.5\,\Omega^{-1}$, the magnetic field is amplified again, provoking a second outflow and the cycle repeats for ever. \\

Our simulation shows clearly that the disc and its periodic outflows
are driven by the recurrent dynamics of the magnetic
field. Off-midplane magnetic pressure 
is responsible for disc compression and assists the launching of a wind near the midplane, whereas the magneto-centrifugal effect accelerates the wind further away. However, the reasons for which the magnetic field evolves into such different configurations remain unclear at this stage and will be investigated in section \ref{investigation}.

\label{cycle_simulations}

\subsection{Dependence of solutions on the vertical field}
\label{Bz_dependence}

We examined how the cycle depends on the strength of the vertical field, which is an important quantity in this problem. Because the cycle of Fig.~\ref{fig_cycle1} is stable in the parameter space studied, it is straightforward to perform its continuation in $\hat{B}_{z}$. As $\hat{B}_{z}$ is slightly decreased (or increased), it takes generally a certain time before the flow converges to a new periodic solution. Figure \ref{fig_cycle_Bz} shows the cycle projections for six different $\hat{B}_{z}$, assuming $\mathbf{\delta=0.033}$ and $z_i=0.5$. The results indicate that the amplitude of the cycle decreases as $\hat{B}_{z}$ decreases. The solution bifurcates to a steady solution (fixed point) for $\hat{B}_{z}\simeq 0.25$.  The average amplitude of the toroidal and poloidal field around which periodic solutions are orbiting increases as $\hat{B}_{z}$ is decreased. This dependence on $\hat{B}_{z}$ is discussed in section \ref{Bz_role}. We found also that the cycle seems to disappear for $\hat{B}_{z}\gtrsim 0.7$, although we failed to explore larger $\hat{B}_{z}$ owing to the increasing computational time in this regime. The state obtained in this highly magnetized disc is non-steady and rather chaotic, which is in contrast with the steady flow obtained by \citep{ogilvie12} for $\hat{B_z}>1$.\\

{Figure \ref{fig_cycle_Bz2} (top left) shows that for $\delta=0.033$, the average mass loss rate does not vary significantly with $\hat{B}_{z}$. In case of $\hat{B}_{z}<0.2$, we found that a new branch of cyclic solutions emerges
after the bifurcation to a steady solution. Their mass-loss rates are similar to the ones
obtained for larger $\hat{B}_{z}$, $\dot{m}_w\sim 0.04$, We
found that the cycle persists for small $\hat{B}_{z}=0.006$ (high $\beta_z=2.8\times 10^4$), but the mass loss rate is significantly weaker $\dot{m}_w\sim 0.01$. Note that the
cyclic solutions obtained for such small $\hat{B}_{z}$ are characterized by a very
slow convergence towards a periodic solution. The transient state is more chaotic and shows
episodic bursts of ejection, each of them having  different mass-loss rates. Fig.~\ref{fig_cycle_Bz2} (top right) reveals also that
the cycle period increases significantly when  $\hat{B}_{z}$ is
decreased. It is approximatively
multiplied by two when $\hat{B}_{z}$ is divided by four. We checked whether these results hold for the critical case $\delta=0$. Fig.~\ref{fig_cycle_Bz2} (bottom) 
shows that for  $\delta=0$ the maximum and average mass-loss rate do
not vary significantly with $\hat{B}_{z}$, as long as the solution has
not bifurcated. The mass-loss rate reaches a maximum for
$\hat{B}_{z}\simeq 0.24$ while it slowly decreases for larger
$\hat{B}_{z}$. As soon as the cycle bifurcates to a steady wind (for $\hat{B}_{z}=0.16$), the mass-loss rate decreases strongly by an order of magnitude. Similarly to $\delta=0.033$, the period increases as $\hat{B}_{z}$ is decreased.} 

\subsection{Robustness of the cyclic behaviour}
\label{rob}
 
\subsubsection{Mass replenishment}
\label{rob_replenish}
\begin{figure}
\centering
\includegraphics[width=\columnwidth]{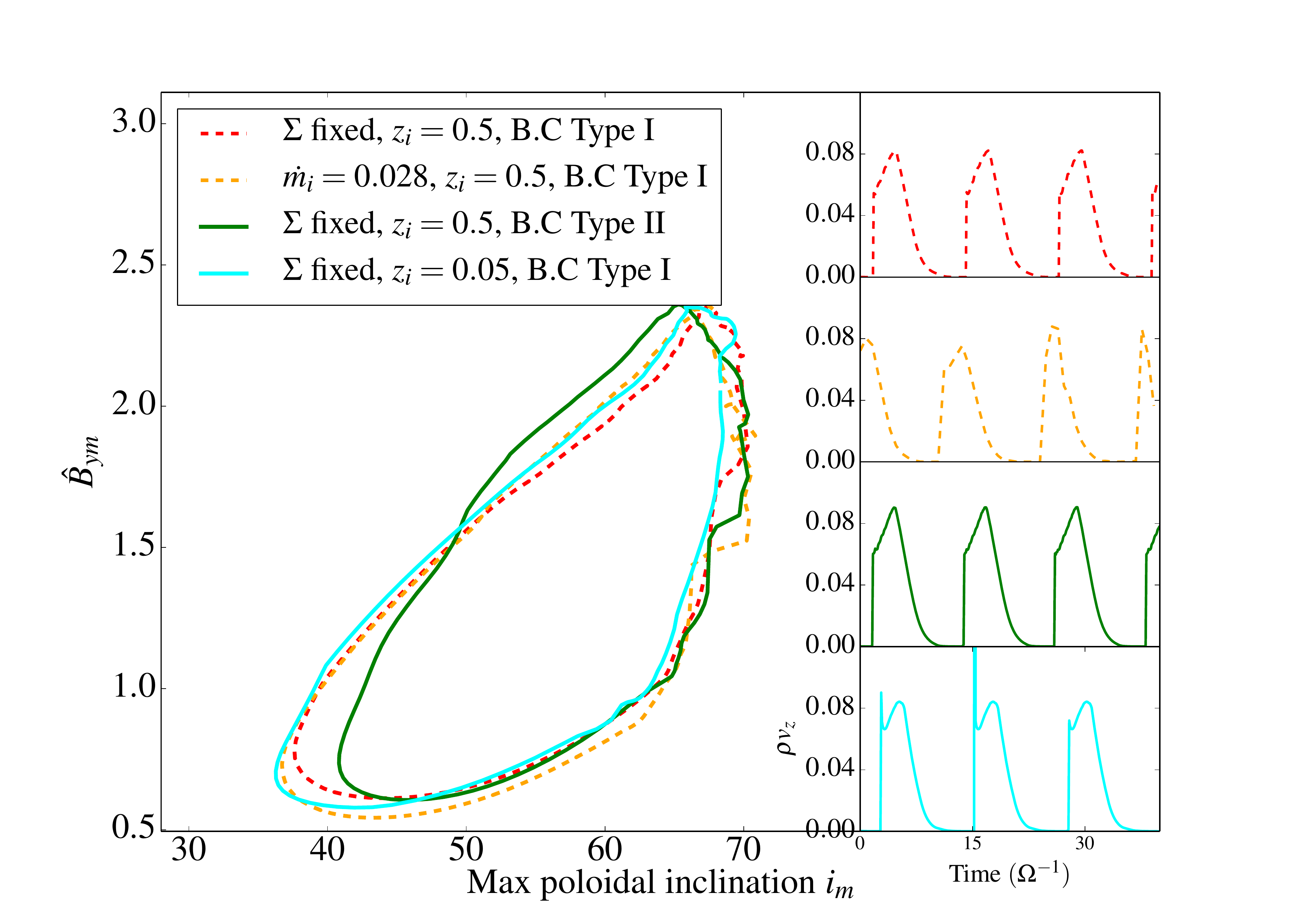}
 \caption{Left: Projection of the cyclic dynamics for in a plane $(i_m,
   \hat{B}_{ym})$ where $i_m$ and $\hat{B}_{ym}$ are respectively the
   maximum poloidal inclination and normalized toroidal field over the
   domain. The red/dashed solution corresponds to the reference cycle
   presented in Fig.~\ref{fig_cycle1} {$\mathbf(\delta=0.033)$}. The yellow/dashed curve
   corresponds to the case of a constant replenishment rate ($\Sigma$
   varying periodically), the solid/green curve to a modified boundary
   condition (type II) while the solid/cyan curve has been computed
   for a lower $z_i$ (altitude below which mass is injected). Right:
   time variation of the mass-loss rate for the different
   cases.}
\label{fig_cycle_bc}
\end{figure}

In the simulations described in \ref{cycle_simulations}, we kept a constant surface density $\Sigma$ by artificially injecting mass near the disc midplane. However, this procedure has one major drawback, as $\varsigma(z,t)$ depends on time and the mass injected near the midplane depends on the properties of the outflow at the upper boundary. If a violent ejection occurs, then a strong injection occurs as soon as the mass has left the box. Replenishing mass in this way by imposing a fixed $\Sigma$ could lead to artificial cycles whose period might increase with the box size. 
To make sure that the cycle we observed is not artificially maintained by the replenishment method, we first checked that the period does not change significantly as $L_z$ is increased or decreased. We also altered the way mass is replenished by assuming a constant mass replenishment with
$\varsigma(z,t)$ given by equation~(\ref{injection}), but with a constant $\dot{m_i}$ whose value is guessed from Fig.~\ref{fig_cycle1} by averaging the mass-loss rate $\dot{m}_w=[\rho {v_z}]_{z=L_z}$ over one period of the cycle.  By using this new replenishment method, we obtained a cycle with very similar properties, shown in Fig.~\ref{fig_cycle_bc} (yellow/dashed curve).
\begin{figure}
\centering
\includegraphics[width=\columnwidth]{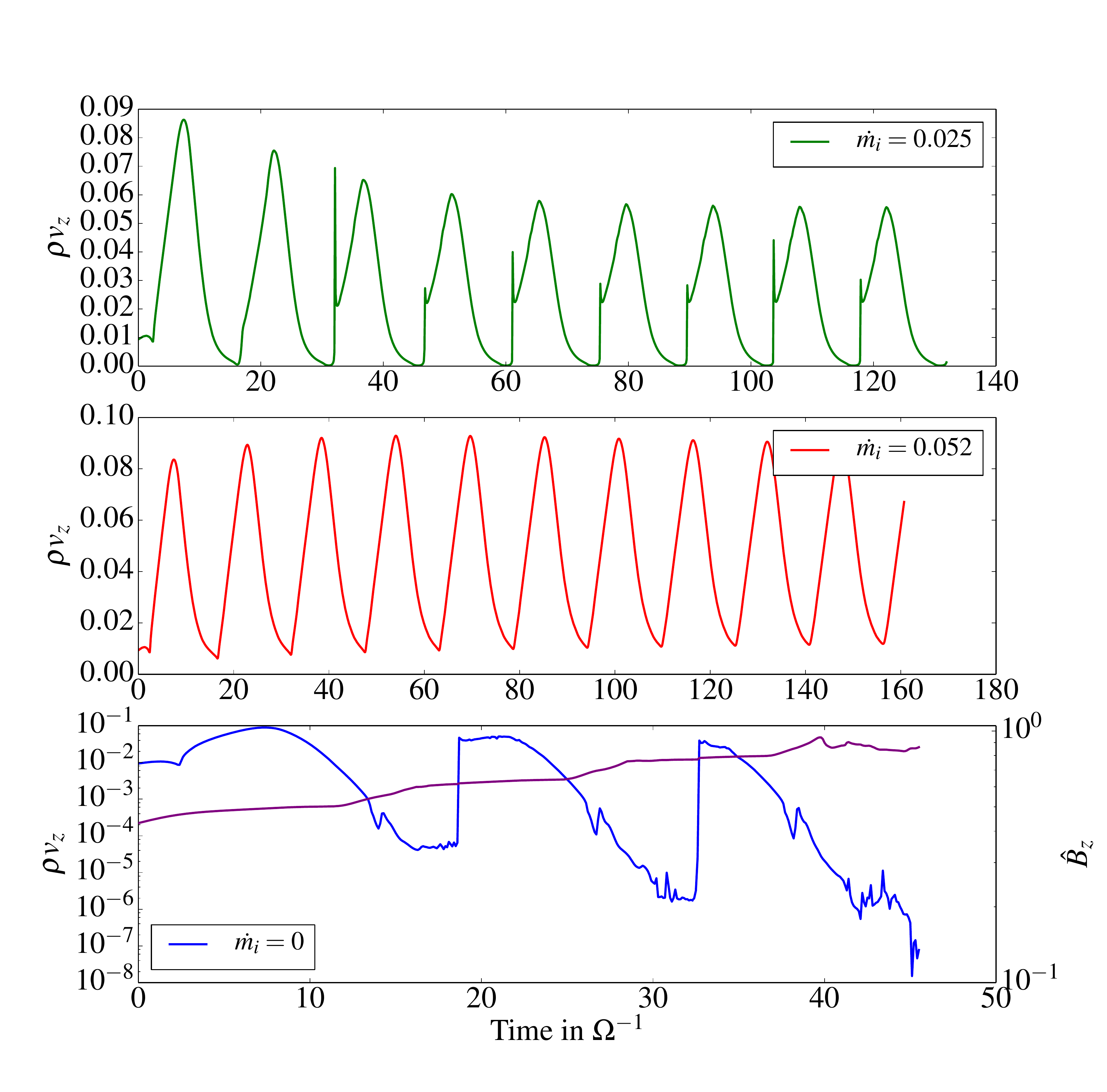}
 \caption{Evolution of the mass-loss rate as a function of time for different mass inflow $\dot{m_i}$. The initial condition is the cycle for $B_z=0.4$, $\dot{m_i}=0.04$ and $\delta=0.033$. In the bottom figure, the purple curve shows the evolution of $\hat{B}_z$.}
\label{fig_cycle_mi}
\end{figure}

One may argue that our replenishment is still artificial. Indeed in a
real astrophysical disc, there is no reason why the mass inflow rate
$\dot{m_i}$ would take such a specific value. It can vary with time
and radius, sometimes independently of the outflow dynamics. We then
ran three simulations with different mass inflow rates $\dot{m}_i=0$,
$\dot{m}_i=0.025$ and  $\dot{m}_i=0.052$, starting from the cycle
computed for $\hat{B}_z=0.4$ and $\delta=0.033$ where
$\dot{m}_w=\dot{m}_i\sim 0.04$ initially. Figure \ref{fig_cycle_mi}
shows that for both $\dot{m}_i=0.025$ and $\dot{m}_i=0.052$, the
initial state relaxes toward a new periodic solution with different
surface density $\Sigma$ (averaged in time). In fact, changing
$\Sigma$ is somehow equivalent to changing $\hat{B}_z$, as $\hat{B}_z$
is inversely proportional to $\sqrt{\Sigma}$. One notable result is
that the system self-organizes and adjusts so that
$\dot{m}_w=\dot{m}_i$. For a more extreme case $\dot{m}_i=0$, bursts
of ejections are still obtained in the first ten orbits but ultimately
they decay into a weak outflow. The magnetization inferred from $\hat{B}_z$ increases but saturates around unity.  This result is reminiscent of the behaviour observed by \citet{Lesur2013} where a transition is observed from a weakly magnetized state with strong outflow to a more strongly magnetized state with very weak outflow. However, in our case, the disc is disrupted after $t=40 \Omega^{-1}$ although the mass has not been entirely depleted.  In summary, a quasi-periodic state can be sustained for a sufficiently long time, independently of the history of the mass inflow,  provided that this inflow is not too small or reduced to zero. \\

We also checked that changing the typical height $z_i$ below which mass is injected does not significantly affect the solution provided that $z_i$ is not much larger than $H$. In Fig.~\ref{fig_cycle_bc} we compare two different values, $z_i=0.05$ (cyan/solid curve) and $z_i=0.5$ (red/dashed curve), and show that differences are small between the two periodic solutions. In the case where $z_i\geq 3H$, the cycle bifurcates to a steady solution. The reason might be that in such a situation there is no need to extract the gas from the disc to launch a wind, as the mass reservoir in the lower atmosphere is directly replenished. Then the dynamics of the midplane region, which we believe to be responsible for the cyclic dynamics, probably becomes less important.  However, having a $z_i$ that differs significantly from the effective scaleheight (typically $H_{\text{eff}} \sim 0.05-0.5 H$ in our simulations) is clearly not realistic as the mass should be mainly replenished in the bulk of the disc. \\

\subsubsection{Boundary conditions and convergence}
\label{rob_bc}

To study the effects of boundary conditions on periodic solutions, we first analyse their dependence on box size. As explained in section \ref{bc_replenish}, the location of the Alfv\'en point might play an important role in MHD wind simulations. If this point leaves the box then the poloidal field inclination is
no longer determined by the Alfvén point crossing condition but ought to be specified through the upper boundary condition as it is determined by global considerations beyond the scale of the box \citep{ogilvie12}. In our setup, is this critical point crucial for obtaining converged solutions? Also, are the cyclic solutions converged with respect to $L_z$ when the aspect ratio of the disc $\delta$ tends to 0? \\

\begin{figure}
\centering
\includegraphics[width=\columnwidth]{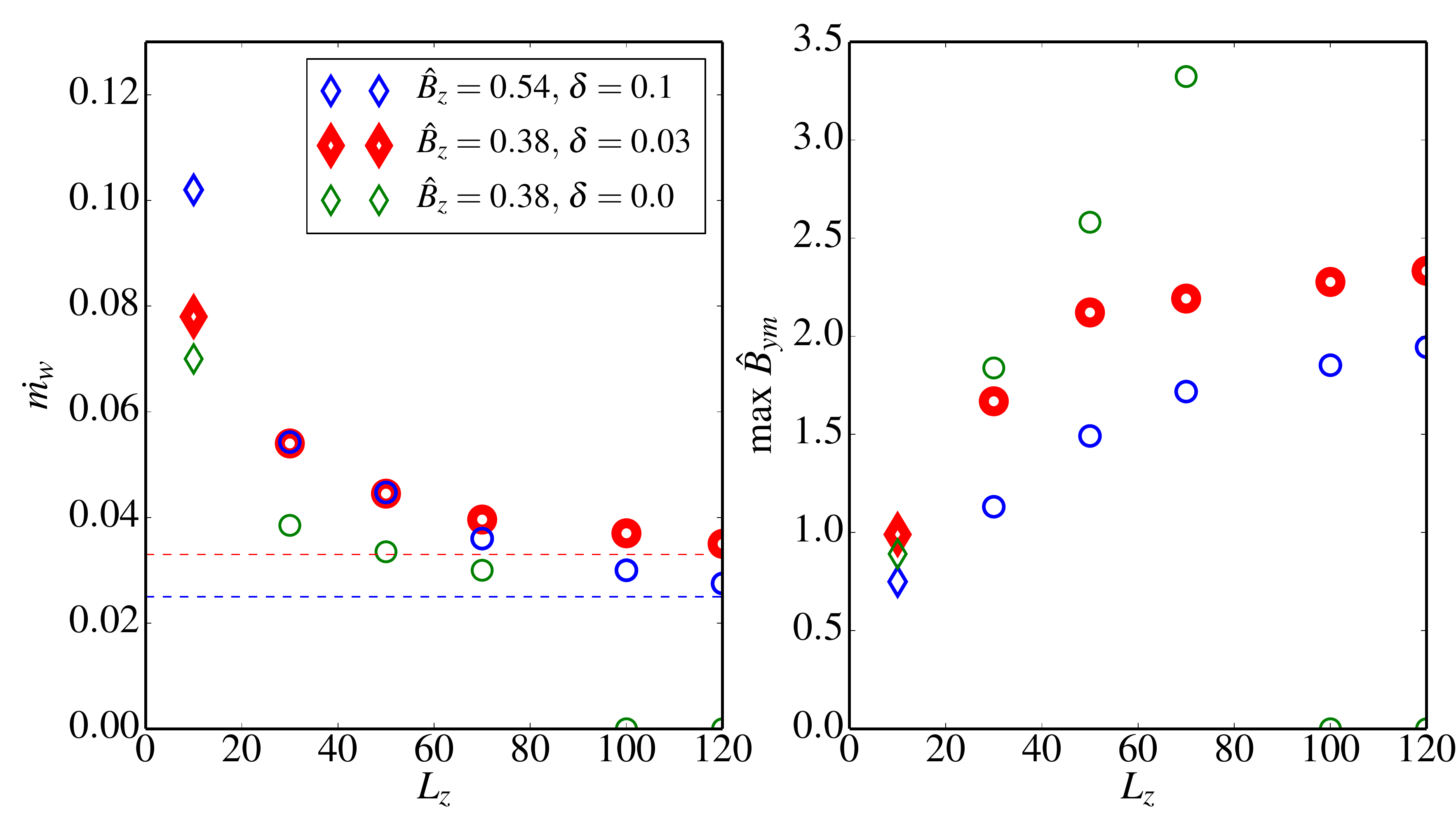}
\includegraphics[width=\columnwidth]{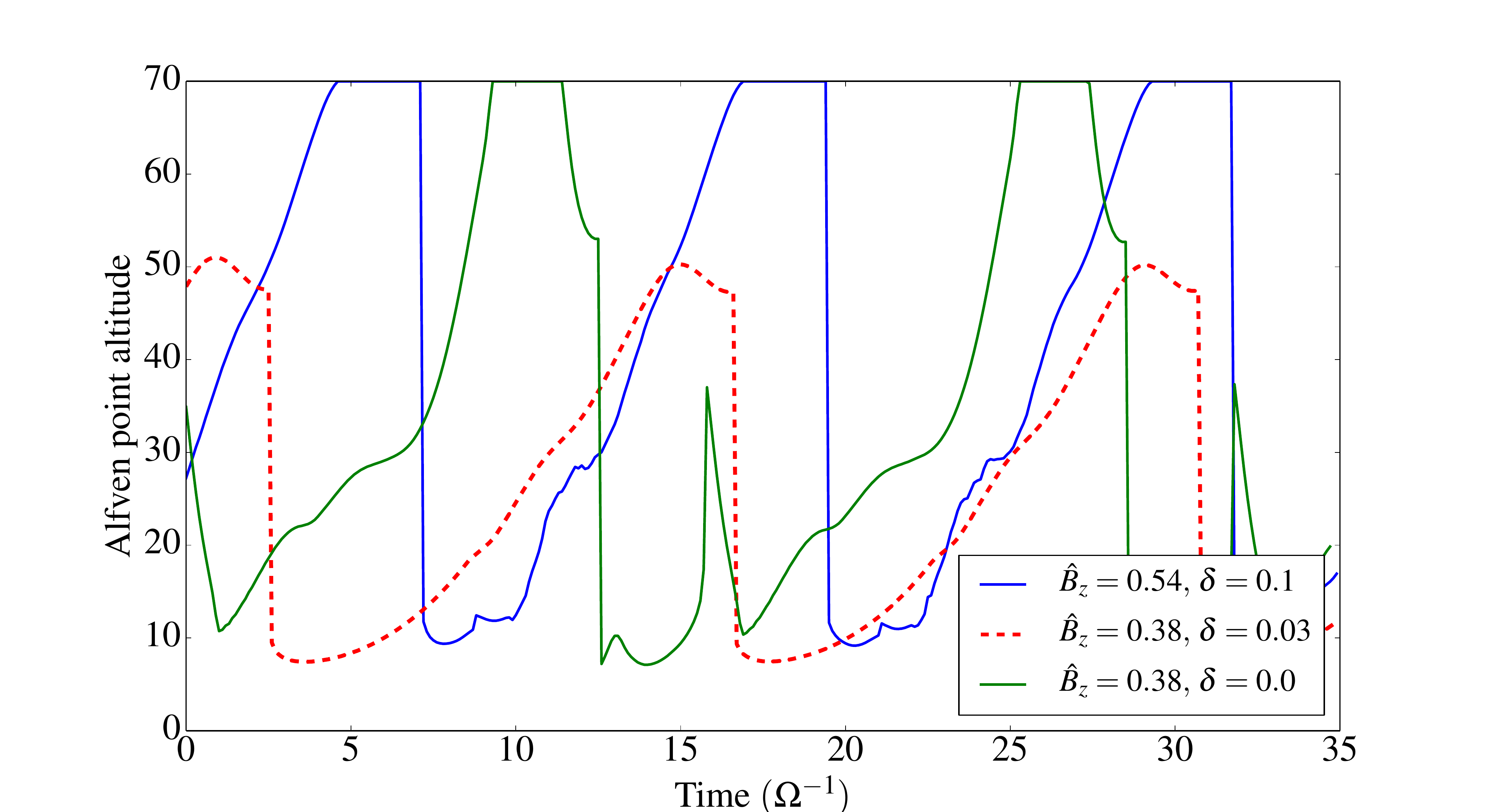}
\caption{Top: average mass-loss rate (left) and maximum of
  $\hat{B}_{ym}$ (right) over one period as a function of the box size
  $L_z$ for three different sets of parameters. Diamond markers
  indicate that the solution has bifurcated to a steady state. Bottom:
  Alfv\'en point altitude as a function of time in these three
  configurations.}
\label{fig_convergence}
\end{figure}

Fig.~\ref{fig_convergence} shows some properties of the reference cycle of Fig.~\ref{fig_cycle1}, such as its average mass-loss rate and the maximum normalized toroidal field $\hat{B}_{ym}$, as a function of $L_z$.  We analysed three different sets of parameters. The first case is $\hat{B}_{z}=0.54$, $\delta=0.1$ for which the Alfvén point lies 80\% of the time inside the computational domain and leaves the box at regular intervals. The second is $\hat{B}_{z}=0.38$, $\delta=0.033$ for which the Alfv\'en point is inside the domain 100\% of the time. Finally the last case corresponds to the limit $\delta=0$ with an Alfv\'en point that is inside the box most of the time. For the two first configurations, the average mass-loss rate $\dot{m_w}$  and $\hat{B}_{ym}$, as well as $i_m$ (not shown here), seem to converge to asymptotic values for large $L_z \gtrsim 70\,H$. In particular the mass-loss rate appears to hover around 0.026 for the first case and 0.035 for the second one. The convergence appears slightly better when the Alfv\'en point stays fully  inside the domain.  Note that the solutions depend strongly on the upper boundary location for $L_z\sim H$, in agreement with previous studies. In the case $\delta=0$, for which the potential diverges as $z\to\infty$, we found that the mass-loss rate starts to converge (but were unable to follow numerically the solution at $L_z>70 \, H$ because of a too strong vertical flow), while $\hat{B}_{ym}$ keeps increasing linearly at large $L_z$. This result suggests that periodic solutions are converged when the box size is much larger than $1/\delta$ {and when the Alfv\'en point remains fully in the computational domain. However, we admit that the convergence is rather slow (one needs to reach high altitudes $z\gtrsim 3 r_0$) and this might be due to the limitation of the shearing box model discussed in section~\ref{box_size_res} and the fact that the fast magnetosonic point is not reached.}\\

We checked also that our cycles are not strongly affected by changing the nature of the upper boundary conditions. Figure \ref{fig_cycle_bc} shows a comparison between the cycle of Fig.~\ref{fig_cycle1} computed with Type I boundary conditions (red/dashed curve), for which the poloidal magnetic field inclination is forced to be 0, and the same solution computed with Type II boundary conditions (solid/green curve), for which the horizontal magnetic field is extrapolated away from the domain. This result suggests that the cyclic mechanism is robust to the conditions imposed by the external medium in the upper atmosphere. 

\subsubsection{Gravity}

\begin{figure}
\centering
\includegraphics[width=\columnwidth]{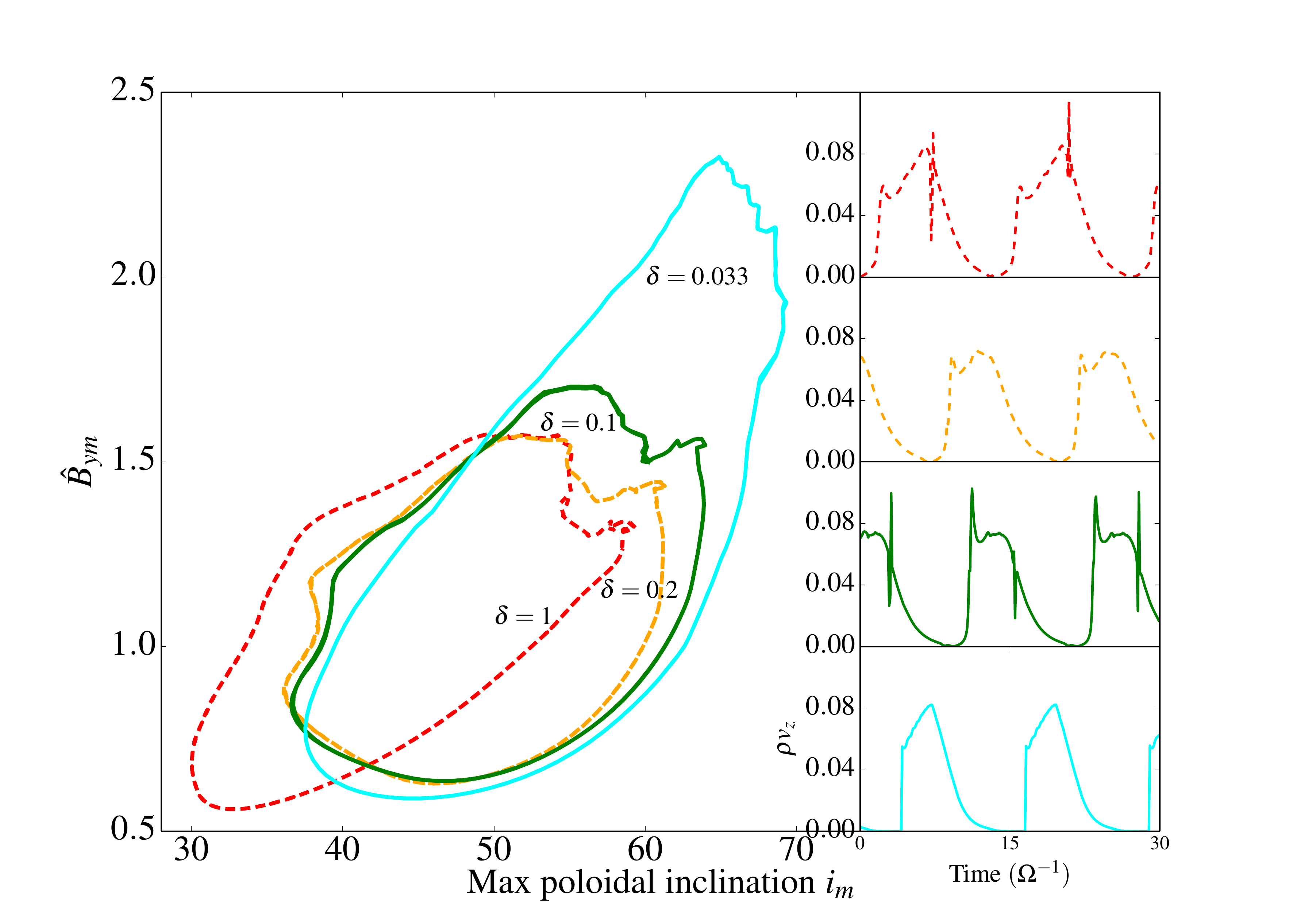}
 \caption{Left: Projection of the cyclic dynamics in a plane $(i_m,\hat{B}_{ym})$. Solutions have been computed for a fixed $\Sigma$, $\hat{B}_{z}=0.54$ and $z_i=0.5$ but with different $\delta$. The red/dashed curve corresponds to the extreme and unrealistic case $\delta=1$. Right: time variation of the mass-loss rate for different $\delta$ (decreasing from top to bottom).}

\label{fig_cycle_r0}
\end{figure}

To understand how the gravity affects the dynamics of these cycles, we performed  several simulations by changing the aspect ratio $\delta$. Simulations are initialized from the reference cycle of Fig.\ref{fig_cycle1} obtained for $\delta=0.033$. We scanned a large range of  $\delta$, going from $\delta=0$ to $\delta=1$.  Figure \ref{fig_cycle_r0} shows that a recurrent dynamics is obtained for all aspect ratios, suggesting that the cycle process is independent of the shape of the gravitational potential at large $z$. {In particular for $\delta\gg H/L_z$ (which correpsonds to $\delta=0.1,\,0.2$ and marginally $\delta=0.033$ for $L_z=70H$) the cycle is only weakly dependent on $\delta$.} However, we note that the amplitude of the magnetic cycle increases as $\delta$ decreases. We obtained also a stronger vertical velocity $v_z$ for $\delta=0.03$ and $\delta=0$.  This is consistent with the fact that a more powerful wind has to be produced to oppose a stronger gravity. A shock propagating upward is also observed when $\delta$ tends to 0, as a result of the strong disc compression. {Note that there are two limits for which the cycle deviates from the original one. First the limit $\delta \rightarrow 0 $. The fact that the solution does not converge in that limit is consistent with the results of Sect.~\ref{rob_bc} which shows that convergence requires $g_z\rightarrow 0$ at large $z$ (the case of a uniformly increasing gravity $g_z$ is clearly problematic).  Second, the limit $\delta=1$ which is inconsistent with the thin-disc approximation and probably irrelevant for astrophysical discs. In that case, the vertical gravity decreases sharply over a scale comparable to $H$, which might change the structure of the disc outflow}.

\subsubsection{Ohmic dissipation}
\label{rob_dissipation}
The simulations described so far have been done in the ideal limit where the effect of microscopic ohmic diffusion is not taken into account. When non-ideal physics is considered, it is possible to define the magnetic Reynolds number \text{Rm} as the typical ratio between the ideal MHD term and the resistive term in the induction equation: 
\begin{equation}
\text{Rm}=\dfrac{3\Omega H^2}{2\eta}.
\end{equation}
We checked that the cycle dynamics is maintained for a large range of Rm $\gtrsim 10$. Protoplanetary discs are generally poorly-ionised but can have magnetic Reynolds number Rm larger than 10, especially in the outer regions. AGN and binary discs have a wide range of resistivity and quite large Rm near their centres \citep{balbus08}.
Therefore we believe that the existence of large-scale periodic outflow solutions in astrophysical discs might not be limited by such dissipative effects, provided that a net magnetic poloidal flux can be maintained for a long time in the disc.

\section{Investigation of the cycle mechanism}
\label{investigation}
\subsection{An MRI-driven cycle?}
\label{MRI?}

We showed in the previous section that the evolution of mass-loss rate
and disc structure during the cycles is intimately connected to that
of the magnetic field. However, the time variability of $\hat{B}_x$
and $\hat{B}_y$ remains unexplained. In order to unveil the nature of
the cyclic process, it seems  necessary to understand the dynamics of
the magnetic field and its interplay with the disc
atmosphere.


As mentioned earlier, the MRI is one obvious candidate to explain the growth of the horizontal magnetic field and the launching of a wind. Some clues, detailed further in section \ref{nonlinear}, suggest that
the MRI is active during the first phase of the cycle. However its nonlinear evolution and connection to periodic behaviour are not obvious as it evolves in a complicated background. In this section, we explore the interplay between the MRI, the stratification, and the wind in order to understand the cyclic magnetic activity. As part of this, it is first necessary to  probe in detail the dynamics of the MRI on a background that is different from the pure hydrostatic
equilibrium, taking into account the strong horizontal magnetic fields
associated with the wind flow.  \\

\subsection{MRI stability of magneto-hydrostatic equilibria}
\label{s:mri}

As the cycle period $T \sim 15 \, \Omega^{-1}$ is significantly longer
than the typical MRI growth time ($\sim\Omega^{-1}$), the dynamics of the midplane region and lower atmosphere can
be modelled to a first approximation as a succession of quasi-magnetohydrostatic
equilibria. The action of the 
MRI is one key ingredient that might control the disc migration from state to state.  We computed a particular class of quasi-steady equilibria, as solutions to suitable boundary value problems, and their stability. Through this study, we aim to identify the conditions for which the MRI is excited and to characterise its behaviour when a background poloidal inclination and a toroidal field are present.  

\subsubsection{Families of 1D nonlinear MHD equilibria}

Nonlinear equilibrium solutions are the local manifestations of the global equilibria calculated by \citet{ogilvie97} and connect to the local winds calculated by \citet{ogilvie12}. They also generalise the weakly nonlinear states
calculated by \citet{liverts12} at the onset of the MRI.  
But being steady, they must
 be distinguished from the exponentially growing channel flow
 solutions of \citet{latter2010}. The wind equilibria are not trivial to obtain and their existence is still matter of debate. To simplify the problem, we first make the assumption that $v_z=0$. This approximation is acceptable for our purposes because the vertical velocity is small compared to the sound speed
in the region where the MRI is active (i.e.\ the disc midplane and the
lower atmosphere). We also do not include resistivity and restrict our study to symmetric equilibria, i.e with $B_x(z=0)=0$ and $B_y(z=0)=0$. Although equilibria also exist in the asymmetric configuration, we found that they take the shape of magnetically `levitating' discs in cases of $\beta_z\sim 1$ and might be 
strongly unstable to buoyancy-type instabilities.\\

The steady nonlinear equations and their solutions are briefly presented in Appendix
B. The system to solve comprises five first-order differential equations and depends on the same number of governing parameters. Two of them are already set by the symmetry. The three remaining parameters can be chosen as the surface density $\Sigma=\int_{-\infty}^\infty \rho \, dz$, the maximum of the normalized toroidal field $\hat{B}_{ym}$ and the maximum poloidal field inclination $i_m$ The two last parameters are directly associated with the quantities we used to describe the magnetic state of cycles. $\hat{B}_{ym}$ is always equal to the asymptotic value taken by $\hat{B}_y$ and quantifies the toroidal magnetic pressure in the upper atmosphere of the disc. Note that although we do not consider any wind in the equilibrium solutions,  
the existence of a toroidal field and its related torque is inevitably
associated with the physical conditions at the outer boundary. 
In a realistic scenario, they can only be maintained by a significant outflow coming from the disc. \\
\subsubsection{Instability in 1D and role of the toroidal field}
\begin{figure}
\centering
\includegraphics[width=\columnwidth]{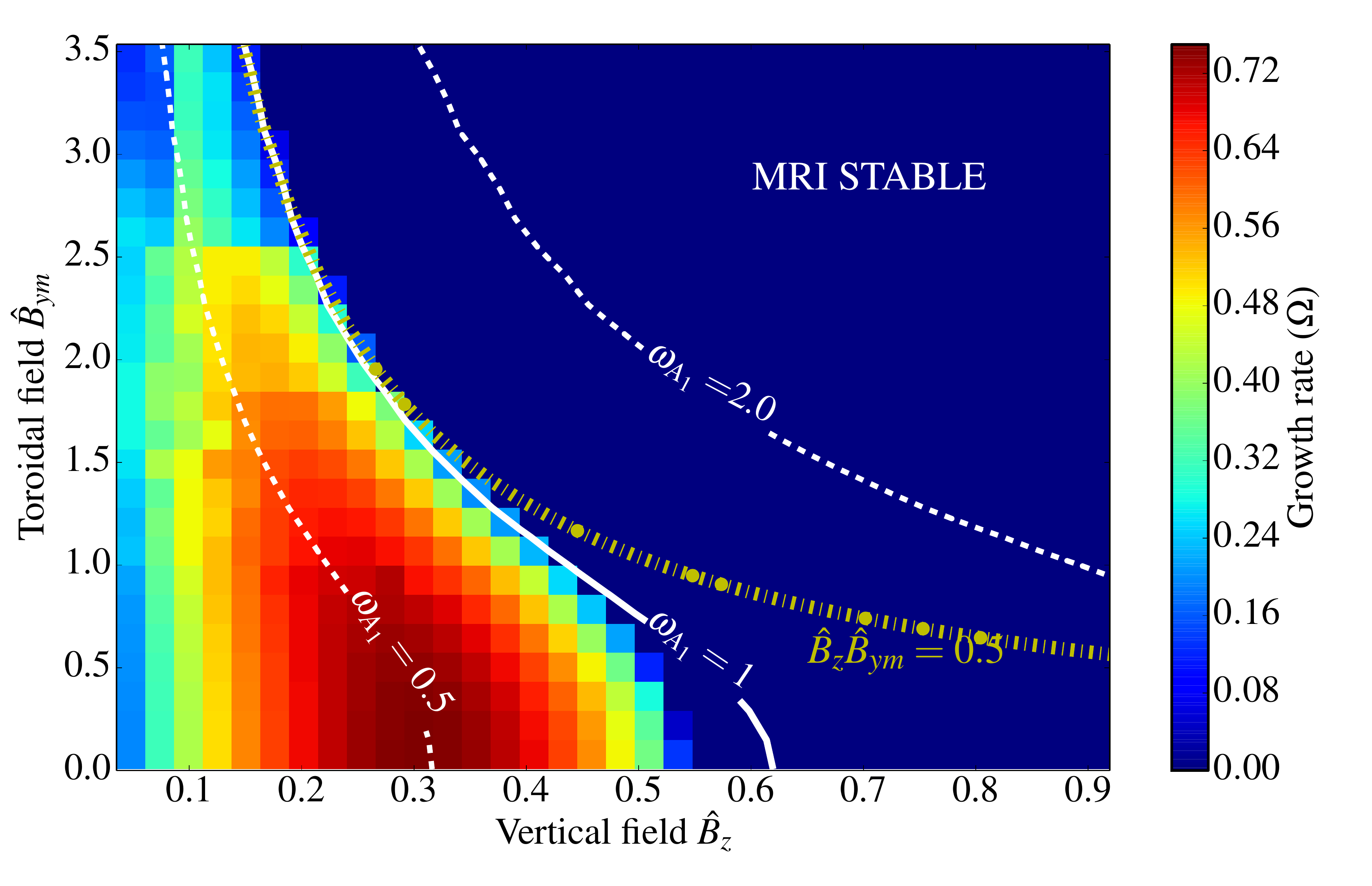}
\includegraphics[width=\columnwidth]{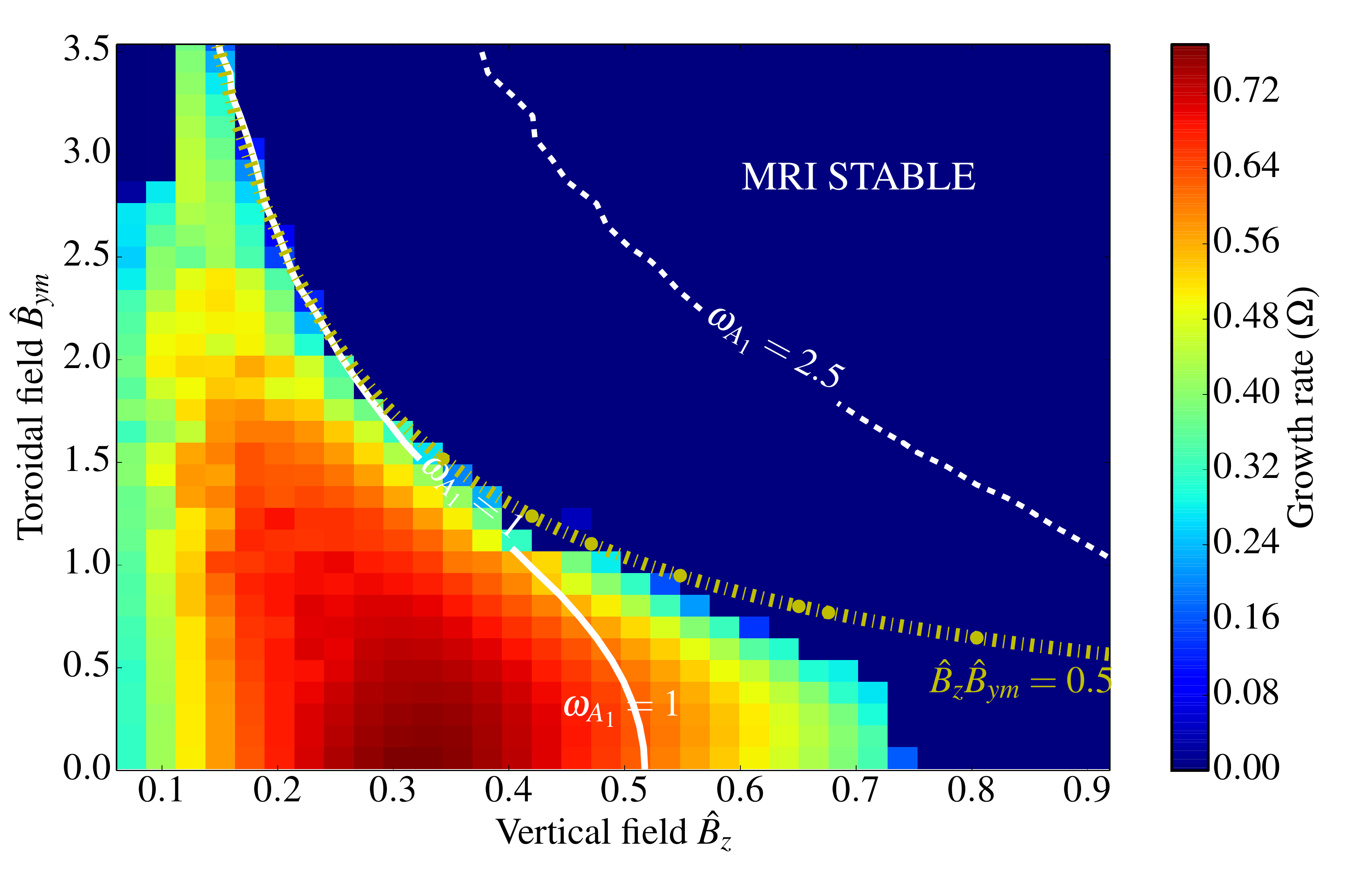}
\caption{Stability of magnetohydrostatic equilibria as a function of
  $\hat{B}_{z}$ and $\hat{B}_{ym}$ for $i_m=0\degree$ (top) and
  $i_m=35\degree$ (bottom). The colourmaps show the MRI growth rates
  of the $n=1$ mode. A blue pixel ($\gamma=0$) indicates a stable
  equilibrium. The white contour lines corresponds to different
  iso-values of the Alfv\'enic angular frequency $\omega_{A_1}$
  defined in equation~(\ref{def_wA}), for which $H_{\text{eff}}$ is obtained
  numerically. The solid line $\omega_{A_1}=1$ seems to lie on the
  marginal stability border of the $n=1$ mode in the case
  $i_m=0\degree$. The yellow dotted line is the marginal stability
  border calculated analytically in the limit $\hat{B}_{ym}\gg 1$ and
  $\hat{B_x}\ll 1$. Note that growth rates were 
 difficult to obtain in the regime of large $\hat{B}_{ym}$ and small
 $\hat{B}_{z}$.}
\label{fig_gammai0}
 \end{figure}

The states computed in the previous subsection constitute exact
steady nonlinear solutions to the governing equations. However, the
dynamical system, if started at (or near) one of these states, may evolve away
because of instability, in particular the MRI but also potentially the
magnetic buoyancy instability \citep{parker66}. In the classical analysis, the MRI occurs when magnetic tension is relatively weak, $\hat{B}_{z}\lesssim 1$, while the buoyancy instability (in isothermal
gas) emerges when the equilibrium is strongly deformed by the poloidal field so that $d\ln(B/\rho)/dz<0$ where $B$ is the norm of the horizontal field \citep{hughes12}. In this subsection, however,  we investigate only the possibility of MRI.  Previous studies in the vertically stratified box have focused
exclusively on the MRI stability of a purely hydrostatic equilibrium
with a uniform vertical field.
\citep{gammie94,latter2010}.
Our results here generalise this to include basic states exhibiting
 strong and non-uniform toroidal 
fields. Some of these results have been computed earlier by 
\citet{ogilvie97,ogilvie98b}, who examined
the stability of discs with
bending poloidal magnetic fields.  Axisymmetric and non-axisymmetric magnetic
instabilities have also been studied in the disc context by \citet{terquem96}.
The stability problem for such equilibria can be solved by numerical methods which are briefly outlined in appendix \ref{appendixB}. 
\\

The three governing parameters are $i_m$, $\hat{B}_{z}$ and
$\hat{B}_{ym}$, which together set the background equilibrium state.
First, for a given inclination $i_m$, we constructed a two-dimensional
stability map in the $(\hat{B}_{z},\,\hat{B}_{ym})$ plane. For each
point in this map, we computed the corresponding equilibria and solved
the linear eigenvalue problem. When the disc is unstable, it is then
possible to report the maximum growth rate of the instability at that
point. Figure \ref{fig_gammai0} shows two of
these maps for different inclinations $i_m=0\degree$ and
$i_m=35\degree$ . Growth rates are computed for the largest-scale
unstable mode which corresponds to $n=1$ in our linearised symmetric
subspace ($n=2$ if the anti-symmetric solutions are taken into
account). The instability is clearly identified as the MRI, since for
$\hat{B}_{ym}\ll1$  growth rates are identical to those obtained by
\citet{latter2010} in the hydrostatic case. When $i_m=0\degree$ (top
figure) and $\hat{B}_{ym}$ is small, the instability reaches its
maximum growth for $\hat{B}_{z}\simeq 0.3$, while it is suppressed
beyond a critical $\hat{B}_{z}\simeq 0.55$. As $\hat{B}_{ym}$
increases, the critical $\hat{B}_{z}$ below which the largest MRI mode
becomes unstable decreases. In other words, for a fixed $\hat{B}_{z}$,
there is a critical $\hat{B}_{ym}$ above which the MRI is
suppressed. The same result is obtained when the poloidal background
field is inclined with respect to the rotation axis. In the case of
$i_m=35\degree$, we found  however that the MRI stabilizes at a 
larger critical $\hat{B}_{z}\simeq 0.75$ for small $\hat{B}_{ym}$ . For both inclinations, we note that the MRI growth rate becomes small, even negligible, when the toroidal field $B_{y_m}\gtrsim 2$.\\

We also computed the same map for larger inclinations, up to
$45\degree$ (not shown here). When the inclination becomes large enough, another
instability seems to arise for $\hat{B}_{z} \gtrsim 1$ and small
$\hat{B}_{ym}$ with a larger growth rate than the MRI. These modes are more difficult to interpret but could be the magnetic buoyancy modes as the background density distribution is strongly deformed by the poloidal field (see the density profile in Fig.~\ref{fig_mhs2} for $i_m =
45\degree$). 

\subsubsection{MRI criterion}
\label{mri_criterion}

A strong magnetic pressure induced by the toroidal
field seems to quench the MRI. This result can be interpreted as
follows: as $\hat{B}_{ym}$ increases, the disc is compressed and its
typical height $H_{\text{eff}}$ is reduced {(with $H_{\text{eff}}$ defined in section~\ref{cycle_pheno})}. This forces the modes 
to be of smaller vertical size, and if this characteristic length is
less than the Alfv\'en length the mode is stabilised. In the classical 
analysis this is equivalent to
 $\omega_A = k_zV_{A_z} >
\sqrt{3}\Omega$. By analogy with the unstratified case, let us define
the Alfv\'enic angular frequency $\omega_{A_m}$ as the product of the
Alfv\'en speed measured in the midplane times the typical wavenumber
$K_m\sim m/H_{\text{eff}}$ of the $n=m$ unstable mode \citep{latter2010}. Then
\begin{equation}
\omega_{A_m}=\dfrac{m}{H_{\text{eff}}}\dfrac{B_z}{\sqrt{\mu_0\rho_0}}=\dfrac{m\hat{B}_{z}}{H_{\text{eff}}}\sqrt{\dfrac{\Sigma c_s\Omega}{\rho_0}}.
\label{def_wA}
\end{equation}
We then expect marginal stability when $\omega_{A_m}\sim 1$ (in our units).
We compute $\omega_{A_1}$ ($n=1$) numerically
for each equilibrium and plot its iso-contours on top of the first
colourmap of Fig.~\ref{fig_gammai0}. For $i_m=0\degree$, it seems
clear that the $\omega_{A_1}=1$ curve lies close to the stability
border of the $n=1$ mode, indicating that $\omega_{A}$ is a suitable
quantity to infer the stability of the MRI. We checked that
$\omega_{A_1}=0.5$ (dotted line in top of Fig.~\ref{fig_gammai0}) or
equivalently  $\omega_{A_2}=1$ lies also on the marginal stability of the $n=2$ mode.
For larger inclinations ($i_m=35\degree$), the $\omega_{A_1}=1$ curve
reproduces the stability border of the $n=1$ mode only for lower
$\hat{B}_{z}$. The discrepancy at stronger vertical fields might issue
from the fact that in this regime,
equilibria display an extended height (see Fig.~\ref{fig_mhs2} of appendix \ref{appendixB}), allowing the MRI mode to be of larger scale and therefore unstable for larger $\hat{B}_z$. Another reason might be that the strong positive vertical shear (a result of
Ferraro's isorotation law), combined with the vertical motions of perturbations, can increase locally the shear rate seen by two fluid particles. \\

We show in Appendix \ref{appendixB} that an analytical expression for
the marginal stability border can be obtained in the limit of large
toroidal field $\hat{B}_{ym}\gg 1$.  Our calculation shows that this
border, drawn in Figure \ref{fig_gammai0} (dotted yellow curve), is
given by $\hat{B}_{ym}\hat{B}_{z}\simeq 0.5$ for the $n=1$ mode in the
symmetric subspace.  It is straightforward to check that this curve
lies along the MRI stability border at large $\hat{B}_{ym}$ for both
inclinations. The theoretical curve fits the numerical stability
border for $\hat{B}_{ym}\gtrsim 0.8$ in the case of moderate 
inclination $i=35\degree$ and for $\hat{B}_{ym}\gtrsim 1.5$ in the case of zero inclination.
\begin{figure}
\centering
\includegraphics[width=\columnwidth]{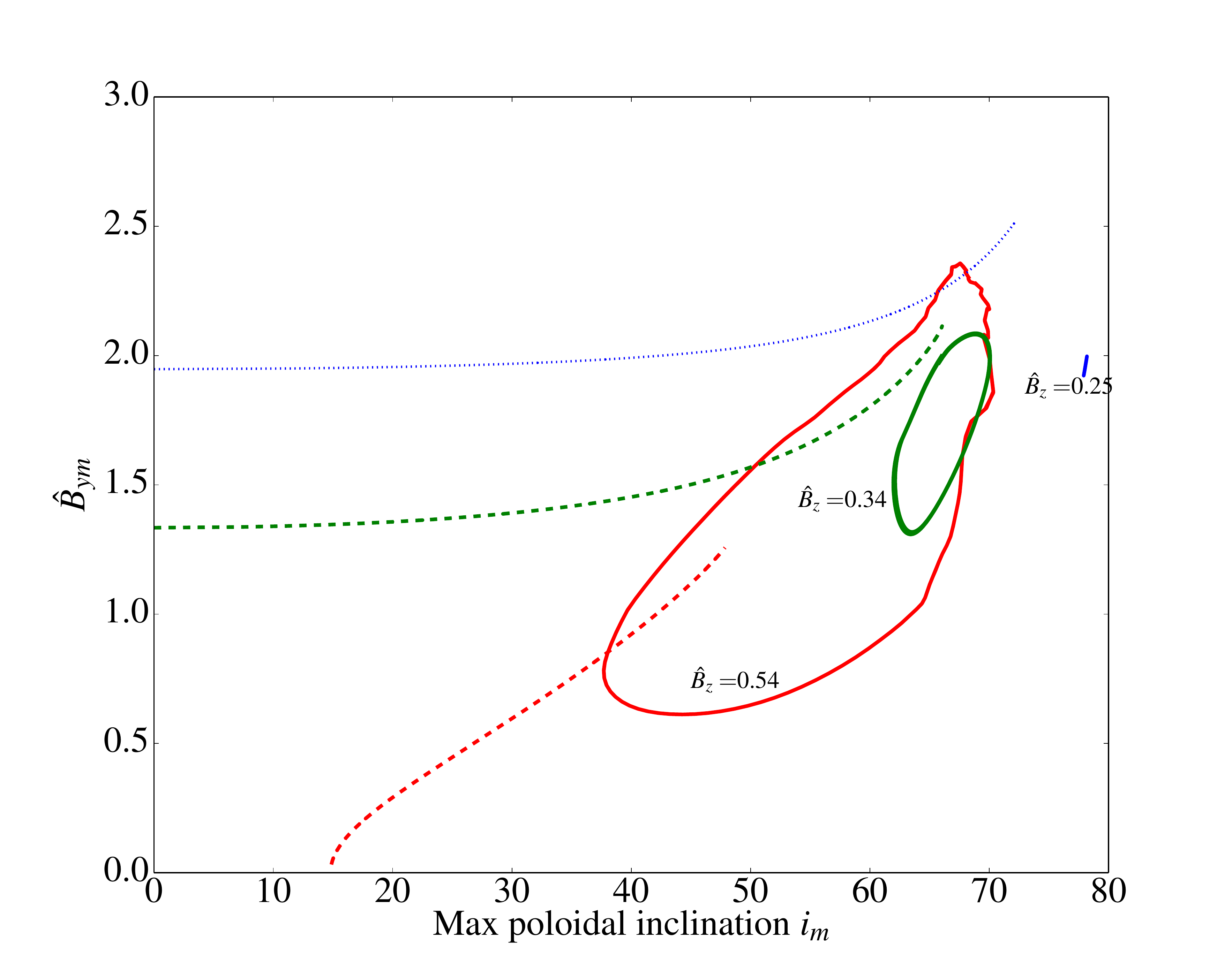}
\caption{Cycles (plain curves) and marginal stability borders of the largest MRI symmetric mode (dashed curves) for different $B_z$. Red, green and blue colors  corresponds  respectively to $B_z=0.54$, $B_z=0.34$ and $B_z=0.25$. Cycles are computed for $\delta=0.033$ and $z_i=0.5$. Note that for $B_z=0.54$, the branch of equilibrium that we follow cannot be continued at larger inclination.}
\label{fig_mristab}
\end{figure}

\subsection{Nonlinear evolution of the MRI and connection with the cycles}
\label{nonlinear}

\begin{figure*}
\centering
\includegraphics[width=0.75\textwidth]{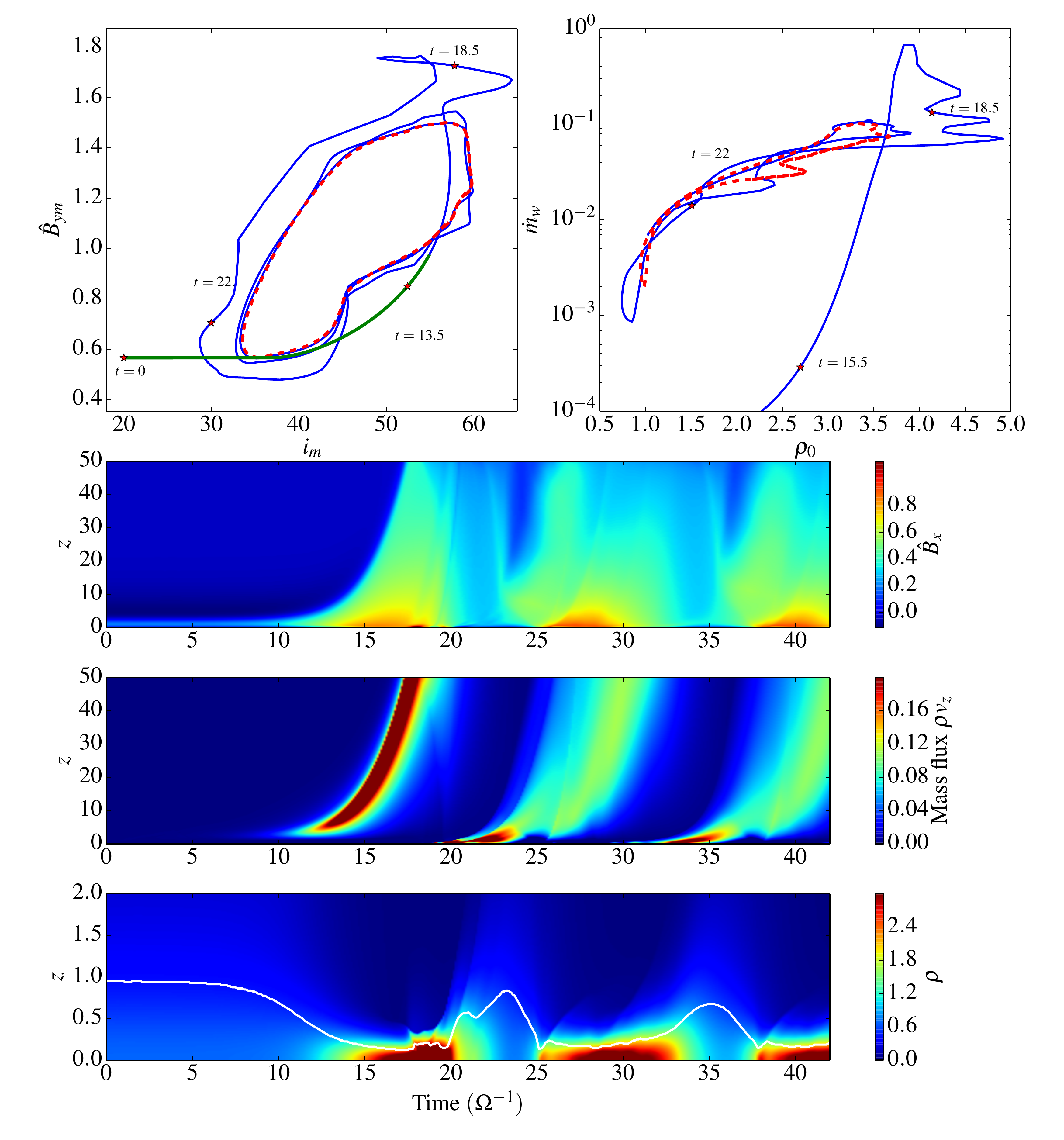}
\caption{Nonlinear evolution of an MRI eigenmode growing on top of a
  magnetohydrostatic equilibrium, for $(\hat{B}_{z}, \hat{B}_{ym},
  i_m)=(0.54, 0.56, 20\degree)$ and $\delta=1$. Top: projection of the
  dynamics in a plane $(i_m, \hat{B}_{ym})$ (left) and in a plane
  $(\rho_0, \dot{m_w})$ (right) where $\rho_0$ is the midplane density
  and $\dot{m_w}$ is the mass-loss rate. Simulation starts at $t=0$
  from the perturbed magnetohydrostatic equilibrium. The green curve
  is the trajectory in phase space that corresponds to the linear
  amplification of the MRI eigenmode. The red dashed curve is the
  periodic solution toward which the dynamics is converging. Bottom:
  Spacetime diagram showing the evolution of $\hat{B}_x$, $\rho v_z$
  and $\rho$. In the last colourmap, the vertical extent is reduced
  to $0\leq z\leq2$ for visibility reasons. The white line corresponds
  to the disc scaleheight $H_{\text{eff}}$.}
\label{fig_nonlinear1}
 \end{figure*}
The aim of this subsection is to link the stability analysis of the equilibria studied in the last section with the cycles found in section \ref{periodic_sol}.
We showed that for a given cycle, a strong toroidal field can build up in the upper atmosphere. The results of section \ref{s:mri} suggest therefore that the MRI is stabilized by such toroidal field at some stage of the cycle. Fig.~\ref{fig_mristab} indicates that for a fixed $\hat{B}_{z}=0.54$, the projection of the reference cycle into a plane $(i_m,\hat{B}_{ym})$ periodically intersects the MRI stability border. This means that it alternates from an MRI-unstable to an MRI-stable configuration. For a lower $\hat{B}_{z}=0.34$ close to the bifurcation point, the cycle becomes tangent to the stability curve while right after the bifurcation, for $\hat{B}_{z}=0.25$, it is completely embedded in the unstable region. This figure indicates that a connection might exist between the cyclic dynamics and the change in stability of MRI channel modes.\\

Although the suppression of the MRI during the second phase of the cycles seems to play a central role in their dynamics, it does not explain  precisely why the disc returns to its weakly magnetized state. To understand more deeply the link between the MRI stabilization and the decay of the magnetic field,  we examined the nonlinear evolution of an MRI mode using PLUTO. We allow a wind to be produced: $v_z \neq 0$. The initial condition consists of an equilibrium in the three-dimensional parameter space $(\hat{B}_{z}=0.54,\hat{B}_{ym}=0.565,i_m=20\degree)$ perturbed in the direction given by its unstable $n=1$ MRI mode whose theoretical growth rate is $\gamma_{th}=0.674\,\Omega$. We used a box of size $L_z=50$, with a uniform grid of 4000 points and $\delta=1$ (this large ratio is unrealistic but allows us to simulate the linear phase of the MRI in a reasonable amount of time). 

The results of the nonlinear simulation are shown in Fig.~\ref{fig_nonlinear1}. First, as expected, between $t=0$ and $t\sim 13.5$ (in units of $\Omega^{-1}$), the MRI mode is linearly amplified as $\hat{B}_x$ and $v_z$ grow exponentially with rate $\gamma_{num}\simeq 0.66\,\Omega$. The projection of the dynamics in a two dimensional plane (top panels) shows that the toroidal component and the mass flux are also growing exponentially during this phase. The magnetic pressure is then enhanced in the lower atmosphere and the disc is compressed, exactly as described for the cycle of section \ref{periodic_sol}. At $t\simeq 18.5$, the poloidal inclination and $\hat{B}_{ym}$ reach a maximum.  This is followed by a very violent mass ejection and the emergence of shocks produced in the bulk of the disc. At $t=20$, the magnetic field amplitude decreases rapidly while the dynamics becomes more gentle and less chaotic. Finally the magnetic field is amplified again and the flow seems to converge toward a stable cycle indicated by a dashed red line. The cycle phenomenology looks like the one described in Fig.~\ref{fig_cycle_r0} for $\delta=1$.\\
\begin{figure}
\centering
\includegraphics[width=\columnwidth]{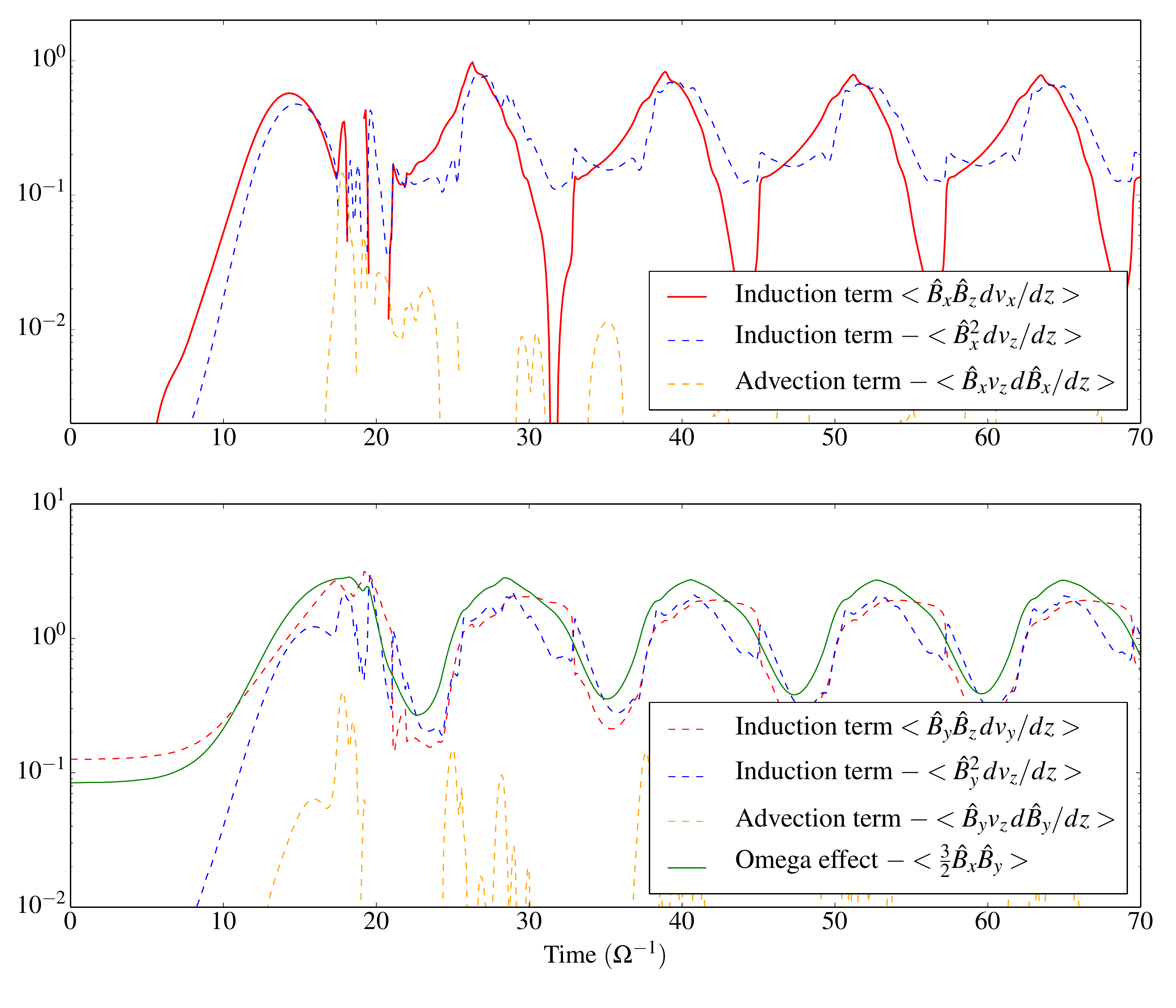}
\caption{Magnetic energy budget for $\hat{B}_x$ (top) and $\hat{B}_y$ (bottom)  as a function of time. Each curve is associated with the
  contribution of a term in the induction equation. Solid lines
  represent positive terms (energy sources) whereas dotted lines
  represent negative terms (energy sinks).}
\label{fig_nonlinear2}
\end{figure}
\begin{figure}
\centering
\includegraphics[width=\columnwidth]{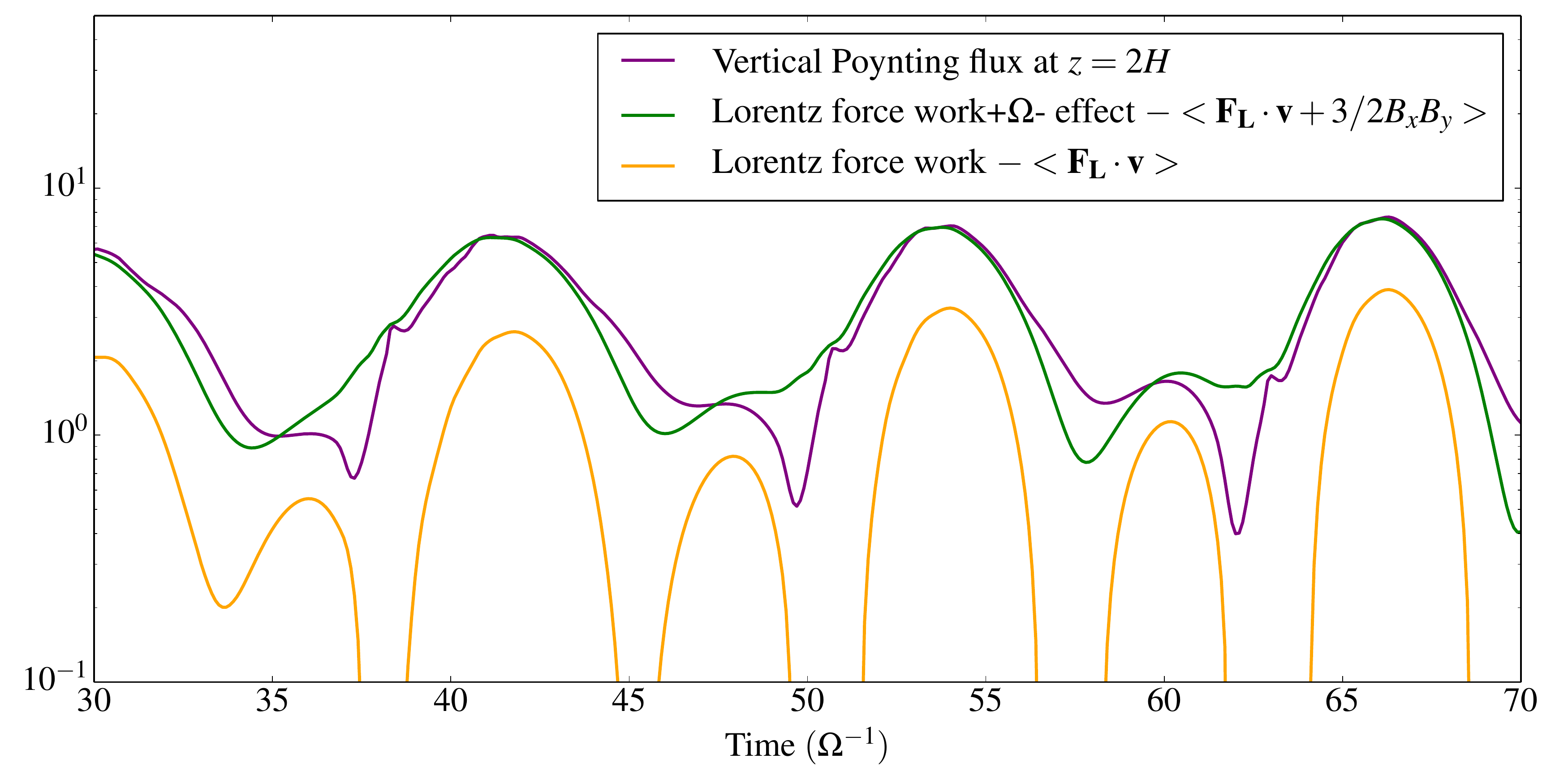}
\caption{Total magnetic energy budget as a function of time. The plain/purple curve is the vertical Poynting flux (outward-directed, so that it is positive when leaving the domain $z\leq 2$) and the green/dashed curve is the integral of $-F_L\cdot\mathbf{v}$ in the domain $z\leq 2$ where $F_L$ is the Lorentz force.}
\label{fig_nonlinear3}
\end{figure}

The periodic evolution of $\mathbf{B}$ and the role of the MRI in generating such cyclic dynamics can be understood by examining the energy budget of the poloidal and toroidal fields. {It is derived from the induction equation and reads:} 
\begin{equation}
\dfrac{\partial}{\partial t}\left(\dfrac{\hat{B}_x^2}{2}\right)=\hat{B}_x\hat{B}_z\dfrac{dv_x}{dz}-\hat{B}_x^2\dfrac{dv_z}{dz}-\hat{B}_xv_z\dfrac{d\hat{B}_x}{dz}
\end{equation}
\begin{equation}
\dfrac{\partial}{\partial t}\left(\dfrac{\hat{B}_y^2}{2}\right)=-\dfrac{3}{2}\Omega \hat{B}_x\hat{B}_y+\hat{B}_y\hat{B}_z\dfrac{dv_y}{dz}-\hat{B}_y^2\dfrac{dv_z}{dz}-\hat{B}_yv_z\dfrac{d\hat{B}_y}{dz}
\end{equation}
{The right-hand side terms in the radial field budget are associated respectively with the linear induction, compressible and advective terms. The contributions to the toroidal field are similar except that the linear induction has an additional term $-(3/2)\Omega B_xB_y$, owing to the shear, also known as the $\Omega$-effect.} 
Fig.~\ref{fig_nonlinear2} shows these different energy contributions integrated in the lower atmosphere from $z=0$ to $z=2H$. During the first phase of the cycle, between $t \simeq 20$ and $t \simeq 26$, $\hat{B}_x$ is amplified and mainly gains its energy from the induction term $\langle \hat{B}_x\hat{B}_z\frac{dv_x}{dz}\rangle$  whereas $\hat{B}_y$ gains its energy from the $\Omega$-effect. This budget is similar to this situation found when the initial MRI eigenmode is linearly amplified (before $t\simeq 13.5$) and is reminiscent of previous budgets of MRI unstable flow \citep{riols2015}. Therefore, this result suggests that the MRI is active during this phase.\\ 

On the contrary, between $t=26$ and $t=34$, when the horizontal components of the magnetic field are maximized, the inductive term associated with the MRI drops sharply and the dominating term becomes $\langle -\hat{B}_x^2\frac{dv_z}{dz}\rangle$. The suppression of the MRI during this phase seems to be directly related to the effect described in section \ref{mri_criterion}. The magnetic compression reduces the disc scaleheight and forces the MRI modes to be of smaller scale. This results in a stabilizing magnetic tension, in the same manner that the MRI is suppressed when the fluid is strongly magnetized. Quantitatively, we have $\hat{B}_{ym} \sim 1.5$ when the disc reaches its maximum compression and the diagram of Fig.~\ref{fig_mristab} indicates that for $\hat{B}_{z}=0.54$ and $\hat{B}_{ym} \sim 1.5 $, the MRI is actually suppressed in the range of inclinations considered here. Note however that we cannot extrapolate directly the marginal stability curves to this nonlinear calculation as they have been computed for $\delta=0$. 

Finally - and this was our main question -  why does the disc return to its initial weakly magnetized state (since we showed that the highly compressed disc is
MRI-stable)? Actually, the second term in the energy equation $\langle -\hat{B}_x^2\frac{dv_z}{dz}\rangle$ is always negative (the advective term being also negative but negligible). Therefore if the MRI is quenched, there is no other source of energy and the magnetic field has to decay. The existence of this negative inductive term, associated with the wind acceleration, clearly breaks the stability of the system and brings it away from its highly compressed state. Note that if the radial field directly loses energy through this term, the toroidal field is affected indirectly
through the $\Omega$-effect. In other words, the wind generated by the MRI and the magnetic pressure gradients in the lower atmosphere have a negative feedback on
$\mathbf{B}$. However, one may wonder how this fraction of magnetic energy is lost physically. Is it transferred to the outflow through Lorentz force, or simply expelled outward by the wind through the expansion of the gas (which differs actually from a pure advective process)? To answer this question, one can re-arrange the induction equation and write the total magnetic energy equation in the form: 
\begin{equation}
\dfrac{\partial}{\partial t}\dfrac{B^2}{2\mu_0}=-\mathbf{\nabla}\cdot \mathbf{S} -F_L\cdot v. 
\end{equation}
Here $S=\mathbf{E}\times \mathbf{B}/\mu_0$ is the Poynting flux which
represents the amount of magnetic energy released by the disc through
a given vertical boundary. $F_L\cdot v$ is the work done by Lorentz
force and quantifies the exchange between kinetic and magnetic
modes. Figure \ref{fig_nonlinear3} shows that during the second phase
of the cycle, when the MRI is suppressed and the magnetic field starts
decreasing,  (between $t=55\Omega^{-1}$ and $t=60\Omega^{-1}$ for
example), the Poynting flux through the boundary $z=2H$ is the
dominating term, indicating that magnetic energy is mainly expelled
outward. At the same time, the term including the Lorentz force work
and the $\Omega$-effect is smaller.  Although the $-F_L\cdot v$ can be
negative, there is no significant transfer from magnetic to kinetic
energy on average. Note that the Lorentz force work is, on the contrary, enhanced during the first phase of the cycle, when the field is amplified by the MRI. By combining the results of Fig.~\ref{fig_nonlinear3} and  Fig.~\ref{fig_nonlinear2}, it becomes clear that the acceleration and the resulting expansion of the gas in the lower atmosphere (owing to the wind) releases the surplus of magnetic energy accumulated in the disc by the MRI and allows a cycle to exist. The same process occurs in a steady wind, but in that case the magnetic energy is continually produced by the MRI and released by the wind without accumulating. 
\\

A similar simulation has been achieved at much lower $\hat{B}_{z}=0.035$. An equilibrium has been selected in this regime and we perturbed it along its $n=3$ unstable eignmode. We used it as an initial condition in a box of $L_z=20$. Again, a cycle was found in the nonlinear phase, with a period slightly longer than that obtained for $\hat{B}_{z}=0.54$. Surprisingly, we found that the average mass-loss rate over one period is $\dot{m}_w=0.08$, larger than the one obtained for higher $\hat{B}_z$. We also found a similar cycle at $\hat{B}_{z}=0.06$ with a completely different code (developed by G.~I.~Ogilvie, which uses high-order finite differences) where $\dot{m}_w\sim 0.1$.  These  two solutions seem to be different from the branch of solutions described in Fig.~\ref{fig_cycle_Bz2}, which undergoes a bifurcation to a fixed point at $\hat{B}_{z}\gg 0.15$. This result shows that powerful cyclic outflow solutions can be found in a large range of $\hat{B}_{z}$ and can involve small-scale modes when the vertical field becomes weak. 

\subsection{Vertical field and condition of existence}
\label{Bz_role}

We showed numerically that the reference cycle studied in section \ref{cycle_pheno} exists above a certain critical $\hat{B}_{z}$. At this critical value, it bifurcates to a steady wind that clearly lies in the MRI-unstable region.  Figure \ref{fig_gammai0} shows that the growth rate associated with the largest-scale MRI mode decreases sharply near this value of $\hat{B}_{z}$, independently of the inclination of the poloidal field.  In addition, Fig.~\ref{fig_cycle_Bz2} seems to indicate that the maximum mass-loss rate is obtained for $\hat{B}_{z}$ close to 0.3--0.4, which corresponds to the maximum growth rate of the MRI mode according to Fig.~\ref{fig_gammai0}. These two results together suggest that the mechanism involved in the periodic behaviour and summarized by Fig.~\ref{fig_process} is sensitive to the growth rate of the MRI mode that drives the cyclic dynamics. In particular, if the growth rate is too small, the wind has time to develop and release the surplus of magnetic energy produced by the MRI. Therefore the system does not develop a strong toroidal field, nor a powerful jet, but still ejects continuously a small amount of mass.   \\

Although the cycle disappears for $\hat{B}_{z} \simeq 0.15$--$0.25$ in the case of $\delta=0$ and for a small range of $\hat{B}_{z}$ for $\delta=0.03$, other periodic solutions seem to exist at much smaller $\hat{B}_{z}$  but are supported by smaller-scale MRI modes (see end of section \ref{nonlinear}). Indeed, the only way to  maintain a significant growth rate in that regime is to excite smaller-scale structures. It is then possible to imagine a family of cyclic solutions for which the dominant MRI mode cascades to small scales when $\hat{B}_{z}$ is reduced. The minimum $\hat{B}_{z}$ that allows a cyclic dynamics will be probably fixed by the dissipative scale. {Also, one could expect even more complex 1D structures than explored so far},  involving the excitation or interaction of multiple MRI modes, as suggested by the long transient chaotic dynamics obtained at small $\hat B_z$ before converging to cyclic solutions.  {However, it is important to emphasize that the cycles or more chaotic 1D structures found in the weak field regime (high $\beta_z$) might be highly idealized  and probably never seen in the fully 3D turbulent discs. This is discussed in more detail in Sect.~\ref{simu_connection}}

\section{Discussion and astrophysical implications}

\subsection{Summary of the results}

\label{cycle_mechanism}
By using one-dimensional isothermal simulations in the local (shearing
box) model of an accretion disc, we have shown the existence of a rich
set of solutions: cycles, steady states, chaotic magnetically
dominated states, transient bursts of ejections. We examined in
particular periodic wind solutions that appear when the disc contains
a vertical magnetic field with midplane plasma beta $\beta_0\gtrsim
1$. The periods of these cyclic solutions are around $10-20$
$\Omega^{-1}$ and seem to decrease with increasing field strength
$B_z$. They are characterized by a sudden and violent ejection, during
which a significant fraction of the mass and angular momentum of the
local region of the disc is carried away, followed by a `quiescent' stage.
They appear to be robust with respect to the variation of several physical parameters (disc aspect ratio, resistivity, vertical field strength, mass inflow) and computational details (grid, boundary conditions, mass replenishment method). The magnetorotational instability (MRI) appears to play a crucial role in the dynamics of these solutions. \\

\begin{figure}
\centering
\includegraphics[width=\columnwidth]{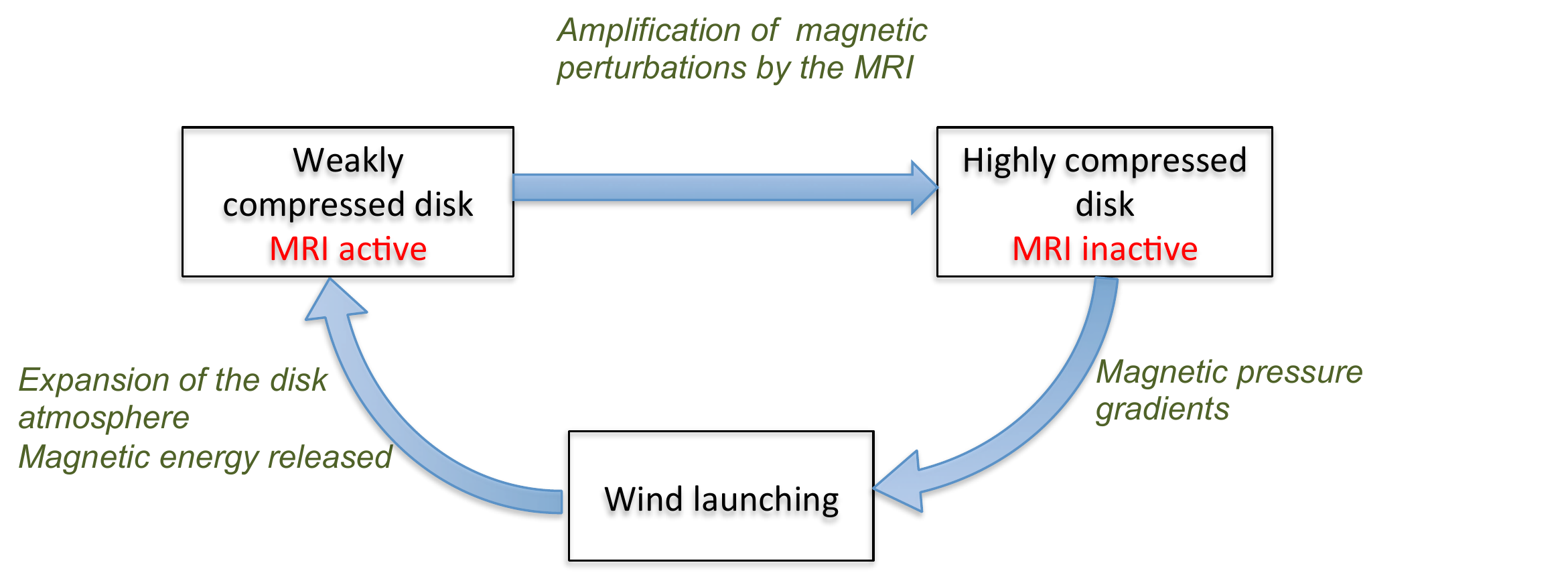}
\caption{A sketch of the possible cycle mechanism}
\label{fig_process}
 \end{figure}

Figure \ref{fig_process} sketches a possible mechanism driving the
cyclic dynamics found in numerical simulations and based on the
different processes discussed in section \ref{nonlinear}. Starting
from a weakly magnetized and compressed disc, the MRI amplifies
horizontal magnetic components, resulting in a strong increase of
magnetic pressure in the lower atmosphere. An outflow is triggered by
magnetic pressure gradients and accelerated by the magnetocentrifugal
effect higher in the atmosphere. At the same time, magnetic pressure
vertically compresses the disc, making the MRI ineffective. The amplification of the horizontal magnetic field stops and the magnetic flux accumulated in the lower atmosphere is expelled outward by the expansion associated with the wind. The first phase of the cycle is then characterized by an MRI amplification whereas the second phase is dominated by the wind dynamics. If the role of the MRI on the wind-launching process has already been highlighted by many authors, we show here that the wind itself, by acting on the large-scale magnetic field, can indirectly affect the conditions of excitation of the MRI. In particular we showed that the amplification of a toroidal field (mainly by the $\Omega$-effect near the midplane) seems to quench the MRI. This feedback makes possible the existence of regular outbursts in the disc atmosphere. 
\subsection{Connections with other MRI-driven wind simulations}
\label{simu_connection}
In this work, we have been able to study disc wind solutions in a particular geometrical configuration (axisymmetric, no radial dependence) and a magnetic regime  where the disc flow is relatively large-scale and cannot be qualified as `turbulent'. The disc is marginally unstable to the MRI so that it oscillates between MRI-unstable and stable configurations. One may ask whether these solutions are relevant or not for astrophysical discs.  

A first point that needs to be discussed is the connection between our
results and the 3D local simulations with very high $\beta_0 \gtrsim
10^{3}$, a regime that has been explored by several authors and for
which the disc is supposed to be fully turbulent. Can we expect
similar periodic structures in that regime with significant mass-loss
rates? The simulations of \citet{fromang13,suzuki09,suzuki10} suggest that a
quasi-periodic outflow can exist when the turbulence is fully
developed. Similarly to our results, the variation of the magnetic
fields is correlated with the evolution of the vertical mass flux,
suggesting that horizontal fields are involved in the cyclic lauching
mechanism.  However their solutions display much weaker oscillations
and mass-loss rates. The orientation of the poloidal field lines also
changes periodically ($B_x$ goes from positive to negative values),
which is not observed in our case\footnote{{A possible
    explanation is that, in our simulations, the expansion of the disc
    associated with the jet is probably not strong enough to remove
    all the energy associated with the MRI mode that triggers this
    jet. Thus, there is no renewal of the MRI seed perturbation, so
    that the polarity and vertical shape of the MRI mode are conserved
    from one cycle period to another.}}. Their simulations also
indicate a more chaotic behaviour, which is expected since many active
modes exist further from criticality. A first step will be probably to
explore 2D axisymmetric simulations in the local or global models with
larger $\beta_0$ to understand whether cyclic solutions exist in this
setup and whether they bifurcate to a more complicated dynamics.  In
particular it would be interesting to see if the cycles develop
instabilities similar to those discussed by \citet{lubow95}, recently
obtained numerically by \citet{moll12} in 2D and \citet{Lesur2013} in
3D.  {In addition, they might be affected or even destroyed by the non-axisymmetric small scale MRI turbulence. Although the cyclic solutions we computed might be unstable or subdominant in the 3D case with small $\hat{B}_{z}$, it is however not excluded that they can form the backbone of a sustained chaotic or turbulent wind dynamics in that regime.} 

{It is also worth comparing our 1D shearing box calculations {with the case of 3D \textit{strongly} magnetised discs ($\beta_0$ comparable to ours), recently simulated by \citet{bai13b} ($\beta_{0_{min}}$=$10^{3}$) and \citet{salvesen16} ($\beta_0\sim 10^{2}-10^{1}$) in the shearing box.} Their simulations show some activity related to the MRI but do not exhibit any cyclic wind motions. It is however possible that their box size (typically $5H$ for one half disc) is not tall enough to capture such periodic wind solutions. {The strongly magnetized simulations are particularly relevant to the so-called “magnetically arrested disks” \citep[MAD, e.g.][]{tchek11,mckinney12} which are thought to exist around black holes. These discs transport a large amount of magnetic flux that accumulates at small radii. Simulations of such discs show strong episodic outflows that release the excess of  magnetic flux at the vicinity of the black hole  \citep{tchek11}. However, it is still unclear how the MRI operates in these objects. It might be quenched as a result of strong magnetic fields. Thus, the cyclic mechanism studied in this work might only be effective during the early phase of the disc, when the magnetic field is building up.\\}

Another issue we addressed in this paper is the dependence of solutions to boundary conditions and in particular the height of the box. Several authors have argued that this effect might result exclusively from the use of the shearing-box model. Actually we showed that one can circumvent this problem by using a modified gravity (which represents the true variation of vertical gravity with height at a fixed cylindrical radius) and ensuring that the height of the box is larger than $H/\delta$, which is the radial distance of the centre of the box from the central object. However it remains unclear in what way a sub-fast magnetosonic outflow depends on the boundary conditions and to what extent the external medium can influence the properties of the outflow. Our preliminary work tends to show that they are independent of the nature of the boundary condition (fixed inclination, extrapolation) if the flow remains super-Alfv\'enic, but more work is needed to understand the role played by the boundaries.  \\\

Finally, it is worth discussing the relevance of shearing-box simulations as they clearly do not reproduce all aspects of a real disc. First, they do not deal with the problem of the evolution of the vertical magnetic flux. How is the large-scale poloidal field maintained during the disc lifetime?  Some studies suggest that the advection of field lines by the accreting disc can be sufficient to accumulate a strong poloidal magnetic field in their center \citep{guilet13}, although these studies do not resolve the MHD turbulence inside the disc, using instead prescriptions for turbulent diffusion. 
Moreover, to simulate periodic solutions, we artificially injected mass near the midplane. However, this mass cannot be computed in a consistent way from the local accretion rate because of the horizontal invariance of our solutions in the shearing box. Global simulations are needed to simulate consistently the accretion-ejection problem and understand how quasi-periodic structures can be obtained in such cases. 

\subsection{Connections with observations}

The idealized setup of the numerical simulations presented in this work (1D local model, isothermal gas) makes it hard to extrapolate directly to astrophysical discs. However, our model might incorporate enough physical ingredients  to allow us to understand some apsects of the interplay between the wind and the MRI dynamics.  First,  there are some observational evidence that real accretion discs (X-ray binaries or AGN) may possess a strong or moderate poloidal field \citep{fender04,miller06,martinvidal15}, which suggests that the periodic behaviour described in this paper is permitted. Second we think that the cycles and the mechanism described in Fig.~\ref{fig_process} might play a role in accretion discs wind dynamics and have a connection with the variability  observed in some protostellar or protoplanetary discs, if the conditions of excitation of the MRI are satisfied in such discs. They could also possibly be involved in the quasi-periodic oscillations (QPOs) measured in the light curves of some discs around white dwarfs, neutron stars or black holes. These phenomena generally have a short period that can be similar to the orbital period $T_0$. For comparison the wind cycles studied here have a period $T \sim 2T_0$. The jet variability measured in our simulations also seems strong enough to be detected in light curves. However, this preliminary work does not allow any firm conclusion on the possible link between a cyclic MHD mechanism and the observed oscillations. A more exhaustive study, comparing instrumental spectral signatures and synthetic data from simulations will have to be achieved to understand if such connections exist.






\bibliographystyle{mnras}
\bibliography{refs} 



\appendix
\section{Existence of steady solutions in the linearised gravity}
\label{appendixA}

We show in this appendix that steady wind solutions with the standard linearised gravity $\hat{g}_z=-\hat{z}$ do not exist. Steady states are obtained by taking ${\partial}/{\partial t}=0$ in equations (\ref{eq_rho})--(\ref{eq_bz}).  Assuming that the viscous term is negligible, we can rewrite equation~(\ref{eq_uz}) as
\begin{equation}
  \rho v_z\frac{dv_z}{dz}=\rho g_z-\frac{dQ}{dz},
\label{dvz}
\end{equation}
where
\begin{equation}
  Q=P+\frac{B_x^2+B_y^2}{2\mu_0}
\end{equation}
is the combined pressure of the gas and the horizontal magnetic field. (The pressure associated with $B_z$ is constant.) For simplicity we assume that $z_i=0$ (mass is replenished only at $z=0$), which implies by mass conservation that $\rho v_z=const=\dot{m_w}>0$. The argument can be generalized to allow for a situation in which all the mass is injected below a certain non-zero height. Integrating equation~(\ref{dvz}) with respect to $z$ from 0 to an arbitrary positive altitude gives
\begin{equation}
\label{eq_constraint}
\dot m_w(v_z-v_{z0})=\int_0^z\rho(z')g_z(z')\,dz'+Q_0-Q,
\end{equation}
where the subscript $0$ refers to values at the midplane.  This equation relates the change in the vertical momentum flux to the weight of the column and the difference in the combined pressure.  Since $\rho$ and $Q$ are positive, while $g_z$ is negative, we deduce that
\begin{equation}
\label{constraint}
v_z<v_{z0}+\frac{Q_0}{\dot{m_w}}.
\end{equation}
(In fact, for symmetrical solutions, $B_{x0}=B_{y0}=0$, so $Q_0=P_0$.) This equation implies that the vertical velocity $v_z$ is bounded above, which in turn means that $\rho$ is bounded below and cannot tend to $0$ at large $z$.  In the standard shearing-box model with $g_z\propto-z$, it is then straightforward to see that the weight integral $\int_0^z\rho(z')g_z(z')\,dz'$ diverges to $-\infty$ as $z\rightarrow\infty$. This makes it impossible to balance equation~(\ref{eq_constraint}) in the limit of large $z$. Therefore steady solutions cannot exist in this case.

In the more realistic modified gravity given by equation~(\ref{eq_gravity}), which has $g_z\propto -z$ for small $z$ but $g_z\propto z^{-2}$ for large $z$, 
the weight integral is finite in the case that both $v_z$ and $\rho$ tend to non-zero constant values as $z\to \infty$.  Therefore steady solutions can exist, in principle, in this case, although they would have an infinite column density.

\section{Nonlinear steady states and numerical methods for                                                                                                                                                                                                                                                                                                                       their stability}
\label{appendixB}

The dynamics of MRI channel modes in a magnetized background can be studied by computing the stability of 1D magneto-hydrostatic equilibria that satisfy the equations (\ref{eq_ux})--(\ref{eq_by}) with $\partial/\partial t=0$ and $v_z=0$. The set of nonlinear equilibria satisfies the following equations:
\begin{equation}
-2\Omega \rho v_y=\dfrac{\partial}{\partial z}\left(\dfrac{B_xB_z}{\mu_0}\right),
\end{equation} 
\begin{equation}
\dfrac{1}{2} \rho\Omega v_x=\dfrac{\partial}{\partial z}\left(\dfrac{B_yB_z}{\mu_0}\right),
\end{equation} 
\begin{equation}
\rho g_z+\dfrac{\partial}{\partial z}\left(-c_s^2\rho -\dfrac{B_x^2+B_y^2}{2\mu_0}\right)=0
\end{equation} 
\begin{equation}
v_x=v_{x0}=const
\label{eq_bx_st}
\end{equation} 
\begin{equation}
-\dfrac{3}{2}\Omega B_x+\dfrac{\partial}{\partial z}\left(v_yB_z\right)=0,
\end{equation}
For a given $\hat{B}_z$, as we are studying solutions that are symmetric about the mid-plane, with $\rho$, $v_x$, $v_y$ being even in $z$ while $B_x$ and $B_y$ are odd, the number of free parameters that describe the whole family is reduced to three. As explained in section~\ref{s:mri}, we can fix the surface density (which is well-defined as long as $\rho$ decays sufficiently rapidly with altitude), the maximum of the toroidal field $\hat{B}_{ym}$ and the maximum inclination $i_m$.

Acceptable solutions of the steady equations require that the radial momentum tends to zero at infinity, which implies that
$B_y$ reaches a constant value at $z=\infty$. It is straightforward to show that the quantity $\hat{B}_{ym}$, defined as the maximum of $-B_y/\sqrt{\mu_0 \Sigma c_s \Omega}$, is directly related to this
asymptotic value. The same argument does not hold for $B_x$ since the poloidal field
inclination $i_\infty$ at $z=\infty$ is not equal to $i_m$. The main reason for choosing $i_m$ as a parameter instead of $i_\infty$  is
that for a fixed inclination $i_\infty$, there is a critical
$\hat{B}_{z}$ below which physical equilibria (with decreasing density
profile) are not permitted. In order to scan the whole parameter
space, it is  then better to use the maximum inclination
(\ref{def_im}), located most of the time near the midplane at $z \leq
H$. \\
All these solutions have a uniform accretion flow $v_{x0}$ in $z$, 
which is driven by the magnetic stress $B_yB_z/\mu_0$ acting 
above the midplane and is directly related to the strength of the toroidal field: 
\begin{equation}
\label{eq_ux0}
v_{x0}=4\,\hat{B}_{z} \,\hat{B}_{ym}\, c_s.
\end{equation}
To compute one of these equilibria, we integrate the system of
differential equations using a Runge--Kutta integrator of 4th order in
space, starting from an estimated midplane density and azimuthal
velocity. These two guessed quantities  are then refined with a Newton
algorithm to match the specified inclination $i_m$ and toroidal field
$\hat{B}_{ym}$. \\
\begin{figure}
\centering
\includegraphics[width=\columnwidth]{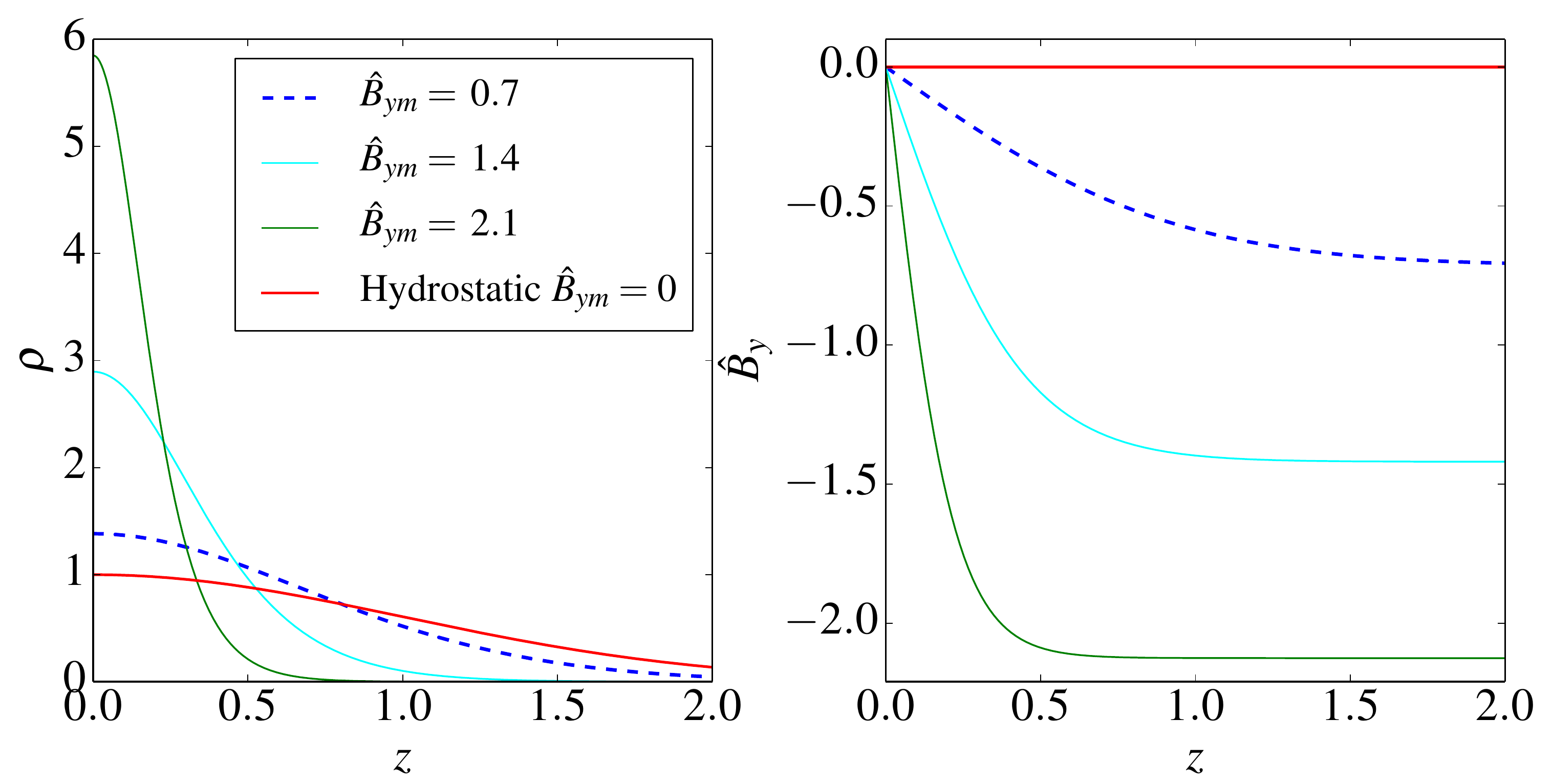}
 \caption{Magnetohydrostatic equilibria for $i_m=0\degree$. Density $\rho$ (left) and toroidal field $\hat{B}_y$ (right) as a function of $z$ are shown for different $\hat{B}_{ym}$. The solid/red curve corresponds to the pure hydrostatic equilibrium.}
\label{fig_mhs1}
 \end{figure}

\begin{figure}
\centering
\includegraphics[width=\columnwidth]{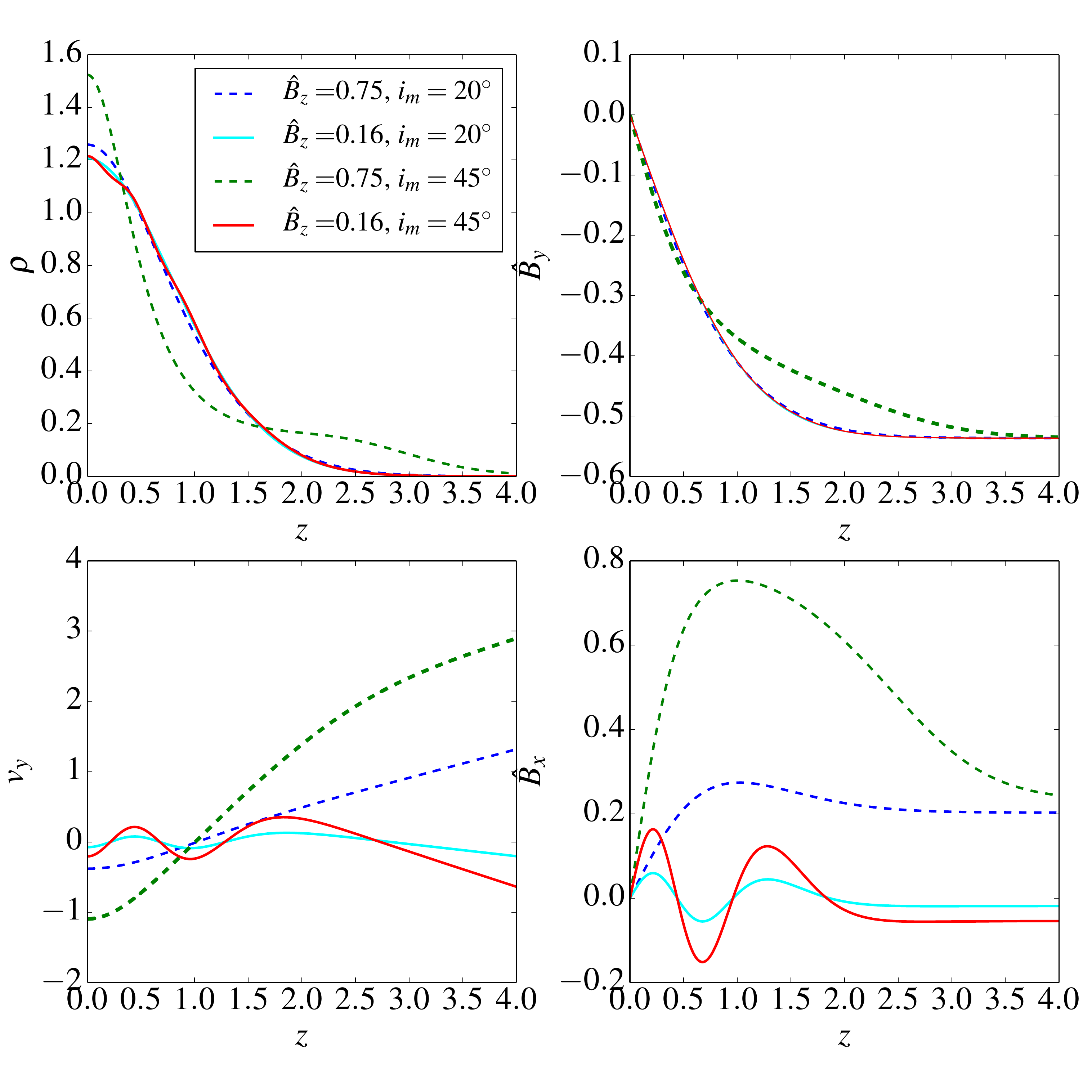}
 \caption{Magnetohydrostatic equilibria for two different inclinations
   $i_m$ and $\hat{B}_{z}$. Quantities $\rho$, $\hat{B}_y$, $v_y$ and
   $\hat{B}_x$ are shown as a function of $z$ (from left to right and
   top to bottom). The toroidal field at infinity is fixed so that
   $\hat{B}_{ym} = 0.53$.}
\label{fig_mhs2}
\end{figure}
Figure \ref{fig_mhs1} shows solutions for $i_m=0\degree$ and three
different $\hat{B}_{ym}$. For this inclination both $B_x$ and $v_y$
are identically zero. 
Increasing $\hat{B}_{ym}$ clearly compresses the disc by reducing
its characteristic vertical scale and by increasing the midplane
density. 
Figure \ref{fig_mhs2}
shows a second set of equilibria computed for inclinations of 
$i_m=20\degree$ and
$i_m=45\degree$.  Increasing $i_m$ gives rise to a
magnetic pressure component localized around $z=H$ which has been induced by the
poloidal field. For large $\hat{B}_{z}$ (dashed curves), this results
in a drastic deformation of the density  profile, since a local
extremum starts developing at large $z$ (this is particularly visible
for $i_m=45\degree$). For small $\hat{B}_{z}$ (solid curves), $v_y$
and $b_x$ oscillate near the midplane with a wavenumber that scales as $\hat{B}_{z}^{-1}$. Note that departure from the
Keplerian flow increases linearly at large distance as a consequence
of field lines co-rotating with the disc and $i_\infty\neq 0$. However
the total $y$ momentum integrated over $z$ remains finite. \\

To study the stability of these equilibria, we added small perturbations on top of them and linearized the system.  Eigenvalues and eigenmodes of the
problem are obtained by using a Chebyshev collocation method on a
Gauss--Lobatto grid. This results in a matrix eigenvalue problem that
can be solved using a QZ method \citep{golub96,boyd01}. Numerical
convergence is guaranteed by comparing eigenvalues for different grid
resolutions and eliminating the spurious ones.  Eigenvalues that are
not symmetric with respect to the imaginary axis are also
eliminated. In the symmetric subspace defined in section \ref{model},
only half of the disc needs to be considered and boundary conditions
for perturbations are imposed at the midplane. Thus $\tilde{b}_x$,
$\tilde{b}_y$, $d(\rho \tilde{v}_x)/dz$ and $d(\rho \tilde{v}_y)/dz$
and $d\tilde{\rho}/dz$ are forced to be 0 in order to be consistent
with the background equilibrium symmetry. We checked that the
eigenfunctions have their momentum $\rho \tilde{v}$ and horizontal
magnetic field components that  tend to 0 as $z$ goes to infinity 
when such boundary conditions are used.

To obtain the marginal stabiliy border of Fig.~\ref{fig_mristab}, we used a different but complementary approach. These curves are obtained  by computing the marginally stable equilibria directly by a shooting method.  The ordinary differential equations governing the vertical structure of a symmetric equilibrium and a symmetric marginally stable linear eigenmode are integrated from an upper boundary (at several scaleheights above the midplane) to the midplane.  At the upper boundary, the horizontal components of the magnetic field are set to specified values, and the density is set to a very small specified value.  The value of $B_y$, and the remaining unknown quantities at the upper boundary, are adjusted by Newton-Raphson iteration so that the symmetry conditions at the midplane are satisfied.
\section{MRI stability in a disc compressed by a strong toroidal field}
\label{appendixB}
An analytical model can be developed for the equilibrium and stability
of a disc that is compressed by a strong toroidal magnetic field
associated with a wind torque.  This corresponds to the limit $\hat
B_{ym}^2\gg1$ of the problem considered in section~\ref{s:mri}.

In a steady state with no outflow, equation~(\ref{eq_uy}) states that
\begin{equation}
  \frac{1}{2}\rho\Omega v_x=\frac{B_z}{\mu_0}\frac{dB_y}{dz},
\end{equation}
while equation~(\ref{eq_bx}) requires $v_x$ to be independent of $z$.  If the disc is compressed mainly by the toroidal field (rather than by gravity or by the radial field), then equation~(\ref{eq_uz}) may be approximated as
\begin{equation}
  \frac{d}{dz}\left(\rho c_s^2+\frac{B_y^2}{2\mu_0}\right)=0.
\end{equation}
The exact solution of these equations is
\begin{equation}
  \rho=\frac{\Sigma}{2h}\text{sech}^2\left(\frac{z}{h}\right),
\label{sech}
\end{equation}
\begin{equation}
  B_y=-B_{ym}\tanh\left(\frac{z}{h}\right),\,\,\,\,\, v_x=-\frac{4B_{ym}B_z}{\mu_0\Sigma\Omega},
\end{equation}
\
where
\begin{equation}
  h=\frac{\mu_0\Sigma c_s^2}{B_{ym}^2}
\end{equation}
is a scaleheight.  Thus
\begin{equation}
  \frac{h}{H}=\frac{\mu_0\Sigma c_s\Omega}{B_{ym}^2}=\frac{1}{\hat B_{ym}^2},
\end{equation}
which should be $\ll1$ for the approximation used above to be valid.  The equivalent value of the effective scaleheight (at which the density has fallen by a factor of $e^{-0.5}$) is $H_{\text{eff}}\approx0.737h$.

The density profile~(\ref{sech}) is the same as that of an isothermal disc compressed by self-gravity (in which case $h=c_s^2/\pi G\Sigma$).  Its stability with respect to channel modes independent of $x$ and $y$, in the presence of a uniform vertical magnetic field, has been determined by \citet{gammie94}. The eigenfunctions for the horizontal displacement are proportional to Legendre polynomials $P_n(\tanh(z/h))$, where the integer $n\geq1$ is a vertical mode number.  The disc is unstable to the MRI if
\begin{equation}
  0<n(n+1)\frac{B_z^2}{\mu_0\Sigma h}<\frac{3}{2}\Omega^2.
\end{equation}
The largest-scale MRI mode within the symmetric subspace considered in
this paper is $n=2$, which is unstable for
\begin{equation}
  0<\hat B_{ym}^2\hat B_z^2<\frac{1}{4}.
\end{equation}
This is equivalent to saying that the MRI is suppressed if the
accretion flow $|v_x|$ produced by the torque associated with the MHD
wind exceeds $2c_s$.


\bsp	
\label{lastpage}
\end{document}